\documentclass[%
superscriptaddress,
preprint,
showpacs,preprintnumbers,
nofootinbib,
 amsmath,amssymb,
 aps,
]{revtex4-1}
\usepackage{graphicx}
\usepackage{dcolumn}
\usepackage{bm}

\usepackage{color}
\graphicspath{ {images/} }
\begin{document}
             
\preprint{JINR report No. E2-2019-41}
             
\title{First-order phase transition from hypernuclear matter to deconfined quark matter obeying new constraints 
from compact star observations}

\author{M. Shahrbaf}
\affiliation{%
 Department of Physics, University of Tehran, P.O.B 14395-547, Tehran, Iran 
}%
\affiliation{%
 Institute for Theoretical Physics, University of Wroclaw, Max Born Pl. 9, 50-204 Wroclaw, Poland}
\author{D. Blaschke}
\affiliation{%
 Institute for Theoretical Physics, University of Wroclaw, Max Born Pl. 9, 50-204 Wroclaw, Poland}
\affiliation{%
 Bogoliubov Laboratory of Theoretical Physics, Joint Institute for Nuclear Research, Joliot-Curie Street 6, 141980 Dubna, Russia}
\affiliation{%
 National Research Nuclear University (MEPhI), Kashirskoe Shosse 31, 115409 Moscow, Russia
}%
\author{A. G.  Grunfeld}
\affiliation{%
 CONICET, Godoy Cruz 2290, (C1425FQB) Buenos Aires, Argentina  
}%
\affiliation{Departamento de  F\'{\i}sica, Comisi\'on Nacional de Energ\'{\i}a At\'omica,
	Av. Libertador 8250, (1429) Buenos Aires, Argentina}
\author{H. R. Moshfegh}
\affiliation{%
 Department of Physics, University of Tehran, P.O.B 14395-547, Tehran, Iran 
}%

\date{\today}

\begin{abstract}
We reconsider the problem of the hyperon puzzle and its suggested solution by quark deconfinement within the two-phase approach to hybrid compact stars with recently obtained hadronic and quark matter equations of state.
For the hadronic phase we employ the hypernuclear equation of state from the lowest order constrained variational method and the quark matter phase is described by a sufficiently stiff equation of state based on a color superconducting nonlocal Nambu-Jona-Lasinio model with constant (model nlNJLA) and with density-dependent (model nlNJLB) parameters. 
We study the model dependence of the phase transition obtained by a Maxwell construction. 
Our study confirms that also with the present set of equations of state quark deconfinement presents a viable 
solution of the hyperon puzzle even for the new constraint on the lower limit of the maximum mass from PSR J0740+6620. 
In this work we provide with model nlNJLB for the first time a hybrid star EoS with an intermediate hypernuclear matter phase between the nuclear and color superconducting quark matter phases, for which the maximum mass of the compact star reaches $2.2~M_\odot$, in accordance with most recent constraints.  
In model nlNJLA such a phase cannot be realised because the phase transition onset is at low densities, before the hyperon threshold density is passed. 
We discuss possible consequences of the hybrid equation of state for the deconfinement phase transition in symmetric matter as it will be probed in future heavy-ion collisions at FAIR, NICA and corresponding energy scan programs at the CERN and RHIC facilities.
\end{abstract}

\maketitle


\section{Introduction}
 
The properties of nuclear matter at supersaturation densities $n >> n_0 \approx 0.16$ fm$^{-3}$ 
are still open questions affecting nuclear physics, particle physics and astrophysics. 
Such a state of matter is realised in the core of a neutron star (NS) and its mechanical properties, encoded in the equation of state (EoS) are uniquely related to the corresponding mass-radius relationship via the Tolman-Oppenheimer-Volkoff equations.
Theoretically, one can infer from mass-radius measurements the NS EoS, see Ref. \cite{Steiner:2010fz} for an early attempt.
Because of the absence of reliable radius measurements, however, we may speak at present only about direct constraints on the EoS from observations of masses and radii of compact stars (CS).
Of particular importance are constraints on the maximum mass as one characteristic feature of the EoS.
The recent measurement of the mass $2.14^{+0.10}_{-0.09}~M_\odot$ for the millisecond pulsar PSR J0740+6620 \cite{Cromartie:2019kug} has renewed the requirement of a sufficient stiffness of the EoS at supersaturation densities. 
On the other hand, the compact star radii should be in accord with the bounds on tidal deformabilities derived from the gravitational waves detected by the LIGO and Virgo Collaboration (LVC) from the inspiral phase of the binary neutron star merger GW170817 \cite{TheLIGOScientific:2017qsa}, for instance $R_{1.6M_\odot}>10.7$ km \cite{Bauswein:2017vtn} 
and $R_{1.4M_\odot}<13.6$ km \cite{Annala:2017llu}. 
These new observational data for masses and radii (compactness) provide more stringent constraints for the behavior of the neutron star EoS. 

One of the key questions concerns the composition of NS interiors.
Already in 1960, still before the discovery of pulsars,  it has
been proposed by Ambartsumyan and Saakyan \cite{Ambartsumyan:1960} that hyperons may occur in the core of a
NS, considering noninteracting particles.
In modern calculations with realistic two-particle and three-particle interactions it was found that the appearance of hyperons softens the EoS to the extent that even the mass of typical binary radio pulsars ($\approx 1.35~M_\odot$) could not be described \cite{Baldo:1998hd,Baldo:1999rq,Shahrbaf:2019wex} (weak hyperon puzzle). 
The more severe "strong" hyperon puzzle consists in the fact that the observational lower bound on the value of the CS maximum mass is nowadays well above $2~M_\odot$\footnote{There are recent works suggesting many-body forces at short-distances based on multi-pomeron exchange which not only improve the description of nuclear saturation properties and elastic nucleus-nucleus scattering data but also allow for a NS mass above $2~M_\odot$ when including hyperons, see  \cite{Yamamoto:2015lwa} and references therein.
In relativistic mean field approaches to hypernuclear matter a severe problem with the "hyperon puzzle" can be avoided since a suitably chosen density dependence of the scalar and vector meanfields together with the repulsive $\phi$ meson meanfield provide a sufficient stiffening to reach CS maximum masses well above $2~M_\odot$; see, e.g., \cite{vanDalen:2014mqa,Maslov:2015msa,Maslov:2015wba,Li:2018qaw} for recent works and \cite{Providencia:2018ywl} for a recent review. 
}. 
\newpage

Besides hypernuclear matter, there are other non-nucleonic forms of matter possible in CS interiors, in particular for the most massive stars with highest central densities, e.g., hadronic matter with $\Delta$ isobars \cite{vanDalen:2014mqa,Drago:2014oja}, dibaryons \cite{Vidana:2017qey} and parity doubled states \cite{Mukherjee:2017jzi,Marczenko:2018jui,Marczenko:2019trv}
as well as deconfined quark matter, which all may occur in the normal but also in the superconducting/superfluid state with pairing gaps in the dispersion relations. 
Moreover, there are meson condensates possible, see \cite{Weber:2004kj} and references therein.
As for the change in the composition of dense NS matter, the question arises whether it will be favorable to first excite heavier hadronic species and only at still higher densities to dissociate them into a state of deconfined quark matter or whether deconfinement can occur at sufficiently low densities to circumvent the occurrence of hyperons.
The latter alternative has been suggested as a possible solution to the weak hyperon puzzle in \cite{Burgio:2002sn,Baldo:2003vx}.
Since an equilibrium phase transition (PT) leads to a softening of the EoS and would have worsened the problem with the insufficient maximum mass, a strongly density-dependent stiffening of the EoS after the transition had to be introduced.
To this end, in \cite{Burgio:2002sn,Baldo:2003vx} a density-dependent bag pressure was suggested but maximum masses stayed still well below $2~M_\odot$.
 
There is an old controversy about whether the observation of a CS with $M>2~M_\odot$ would rule out hybrid stars with quark matter cores and thus would remove the quark deconfinement as a solution of the strong hyperon puzzle.  
However, it has been shown, e.g., in \cite{Klahn:2006iw}, that a sufficiently strong vector meson coupling would allow maximum masses of quark-hadron hybrid stars above $2~M_\odot$ and the occurrence of a diquark condensate can induce a lowering of the deconfinement transition so that indeed problems with the high-density behaviour of hadronic EoS like the hyperon puzzle and the direct Urca problem can be solved. 
Using the same three-flavor color superconducting NJL model \cite{Blaschke:2005uj} as in \cite{Klahn:2006iw} and an extension of the density-dependent relativistic meanfield model DD2 \cite{Typel:2009sy} to include $\Lambda$ hyperons and 
an additional repulsive $\phi$ meson mean field it has been shown in \cite{Lastowiecki:2011hh} that with this setting a CS structure with a hypernuclear shell and color superconducting quark matter in the core under fulfillment of the $2~M_\odot$ mass constraint is possible. 
For this solution it was essential to add a medium dependent bag pressure contribution, see also \cite{Bonanno:2011ch} who used a different version of a relativistic mean field theory with hyperons \cite{Lalazissis:1996rd,Lalazissis:2005de}
in its extension to the hypernuclear sector \cite{Colucci:2013pya}.  

In this work, we use the lowest order constrained variational (LOCV) method to provide an EoS for baryonic matter 
for different asymmetry parameters $x=(\rho_n-\rho_p)/\rho_B$ and hyperon fractions $x_\Lambda=\rho_\Lambda/\rho_B$
\cite{Shahrbaf:2019bef}. 
The LOCV approach allows us to study the PT for arbitrary isospin asymmetries. 
 
We shall consider here the deconfinement to a recently developed stiff EoS for deconfined quark matter 
\cite{Alvarez-Castillo:2018pve,Ayriyan:2018blj}, the generalized nonlocal Nambu-Jona-Lasinio (nlNJL) model, which is a recently-developed approach to describe quark matter including color superconductivity.  
In comparison to the original version of the nlNJL model \cite{GomezDumm:2005hy,Blaschke:2007ri}, we introduce here 
density-dependent coefficients which are chosen such that the results of a recent density-functional approach to quark matter
\cite{Kaltenborn:2017hus} can be reproduced.
 
The PT as obtained by a Maxwell construction predicts a jump in energy density which, depending on its size, may even lead to an unstable branch in the CS mass-radius diagram which eventually is followed by a third family branch \cite{Gerlach:1968zz} of stable hybrid stars, disconnected from the second family of pure neutron or moderate hybrid stars \cite{Alford:2013aca}. 
As an observable feature the third family leads to the mass twin phenomenon \cite{Glendenning:1998ag} which recently has
been shown to be possible also for high-mass pulsars \cite{Blaschke:2013ana,Benic:2014jia,Alvarez-Castillo:2017qki}. 
We will employ the interpolation method on the basis of the nlNJL EoS \cite{Alvarez-Castillo:2018pve} which is a powerful and flexible tool to construct a strong PT and to answer the question of the possible existence of a third family of CSs which would require a strong PT in dense matter.
 
In Sec. II we present the theoretical formulation based on the LOCV method and nlNJL model. The Sec. III is devoted to the results and discussion for the properties of hypernuclear matter and hadron-quark matter PT. 
In Sec. IV the properties of PT for isospin-symmetric matter as well as a comparative study on model dependence of PT are presented.  
This includes a discussion of the PT onset in symmetric matter that would follow from the hybrid EoS model constrained by recent NS observations.
Finally, the summary and conclusion are given in Sec. V.

\section{Hybrid star EoS with hyperons and quark deconfinement}
In this work we use a two-phase description in order to construct a transition from hadronic to the quark phase. The theoretical approaches used to calculate the EoS for each of these two phases will be discussed in the following two subsections.

\subsection{Hadronic phase: Hypernuclear matter within the LOCV method}

For the nuclear matter phase, we use a microscopic potential-based technique called LOCV method by Owen et al. \cite{Owen:1977uun} for calculation of the bulk properties of nuclear fluids, such as the saturation properties. Different types of nucleon-nucleon interactions have been employed so far such as Reid68 and $\Delta$-Reid \cite{Modarres:1979jk}, UV14, AV14 and AV18 \cite{Bordbar:1998xv,Modarres:1998ma,Modarres:2000nk}, and charge-dependent Reid potential (Reid$93$) \cite{Moshfegh:2005rom,Moshfegh:2007mxh} while a central potential \cite{Hiyama:2006xv,Hiyama:2002yj} has been used for nucleon-hyperon and hyperon-hyperon interactions recently. 
Since the three body interactions have an essential role to describe the nuclear matter properties, the LOCV method is capable to deal with the three body interactions like Urbana type \cite{Goudarzi:2015dax} and chiral three nucleon force \cite{Goudarzi:2019orb}. 
This method has been used not only at zero but also at finite temperature for the calculation of thermodynamic properties of 
hot and cold fermionic fluids \cite{Moshfegh:2007mxh,Bordbar:1998xv,Modarres:1998ma,Modarres:2000nk,modarres2009,modarres2003}. 
In other variational methods such as the APR method \cite{Akmal:1998cf}, while the calculations are similar to the ones in our method and there is a good agreement between our results \cite{moshfegh2010}, there are some differences in calculating the correlation functions.
Usually, in other variational methods, several free parameters in the correlation functions are chosen so that the energy per nucleon at every given density to be minimized but they do not mention the normalization constraint and therefore it is not clear how the normalization constraint is satisfied. 
In the LOCV method, we impose the normalization condition by demanding the control parameter $\chi$ to vanish,
 \begin{equation}
 \label{eq:1}
 \chi=\left\langle \Psi|\Psi\right\rangle - 1
=\frac{1}{A} \sum _{ij}\left\langle ij|F_{p}^{2}-f^{2}|ij-ji\right\rangle  =0,
\end{equation}
where $F_{p}$ is the Pauli function. For asymmetric nuclear matter, it is defined by
\[ F_{p}(r) = \begin{cases} 
          {\normalsize[1-\frac{9}{2} (\frac{J_{1} (k_{f_{i}} r)}{k_{f_{i}} r} )^{2} ]^{-\frac{1}{2} }} &  {\rm indistinguishable ~particles,}\\
          1 &  {\rm distinguishable~particles}, 
       \end{cases}
    \]
where $J_{1}(k_{f_{i}} r)$ denotes the spherical Bessel function of order $1$ and $k_{f_i}$ is the Fermi momentum of each particle. The wave function of the system then reads
\begin{equation}
\Psi(1...A)=F(1...A)\Phi(1...A),
\end{equation}
where $\Phi(1...A)$ is the uncorrelated Fermi system wave function (Slater determinant of plane waves) and $F(1...A)$ is the many-body correlation function.
In (2), $f(ij)$ denotes to the two-body state-dependent correlation functions.
In the Jastrow formalism, the two-body correlation functions $f(ij)$ are defined as
\begin{equation}
f(ij)=\sum _{\alpha ,p=1}^{3} f_{\alpha }^{p} (ij )O_{\alpha }^{p} (ij), 
\end{equation}
where $O_{\alpha }^{p}$ is the projection operator which projects on to the $\alpha$ channels, i.e., $\alpha=\lbrace J,L,S,T,T_{z},s\rbrace$ where $s$ is the strangeness number of baryons. For singlet and triplet channels with $J=L$ we choose $p=1$ and for triplet channels with $J=L\pm1$ we set $p=2,3$. The operators $O_{\alpha }^{p}$ are given by
\begin{equation}
\label{eq:4}
O_{\alpha }^{p=1-3} =1,(\frac{2}{3} +\frac{1}{6} S_{12} ),(\frac{1}{3} -\frac{1}{6} S_{12} ) ,
\end{equation} 
where $S_{12} =3(\sigma _{1} .\hat{r})(\sigma _{2} .\hat{r})-\sigma _{1} .\sigma _{2}$ is the spin tensor operator with 
$\sigma _{1}$ and $\sigma _{2}$ describing the spins of nucleons 1 and 2, respectively.
In this model, it is supposed that there is a specific form for the long-range behavior of the correlation functions due to 
an exact functional minimization of the two-body energy with respect to the short-range parts of the correlation functions. 
The constraint will be incorporated only up to a certain distance (the healing distance) where the logarithmic derivative of the correlation function matches that of the Pauli function. 
After the healing distance, the correlation function will be replaced with the Pauli function. 
The condition (\ref{eq:1}) also ensures that the correlations are predominantly of the two-body kind and the higher many-body contributions are small and results in a rapid convergence of the cluster expansion.
In comparison with the mean field theories, we can directly use the realistic interactions for baryons which are phenomenologically obtained from scattering data. 
Of course, since the LOCV method is a non-relativistic method, when considering hyperons, our results for the NS maximum mass  are well below the value required by astrophysical observations, in agreement with previous works based on Brueckner-Hartree-Fock (BHF) approach as well as other non-relativistic potential models \cite{Shahrbaf:2019wex,Baldo:1998hd,Baldo:1999rq}. In a Green function method like BHF, however, the correlation functions cannot be obtained.

We point out the following characteristics of the LOCV method:
\begin{enumerate}
 \item a microscopic method in configuration space which is  purely variational;
 \item simply generalizable to finite temperature;
 \item correlation functions for baryon-baryon interactions as well as structure functions which are important quantities of interest in scattering studies, are obtained directly in our formulation. In fact, the two-body energy in LOCV method is a functional of correlation function. Thus, by a minimization of the energy with respect to the correlation function, one can obtain both of the energy and correlation function; 
 \item tensor correlation functions are employed;
 \item considering Eq.(\ref{eq:4}), the energy per baryon and correlation functions are state-dependent and can be obtained for each state which is defined by $\alpha=\lbrace J,L,S,T,T_{z},s\rbrace$;
 \item numerical calculations are not a time consuming, so that they can be performed on standard desktop or laptop computers.
 \end{enumerate}
Recently, we have extended the work of Goudarzi et al. \cite{Goudarzi:2015dax} by including hyperons, and studied the composition and EoS of a charged neutral, equilibrated mixture of neutrons, protons, electrons, muons, free $\Sigma^{-}$  as well as $\Lambda$ hyperons at zero temperature. In this work, the AV18 nucleon-nucleon interaction supplemented with Urbana type three body force \cite{Wiringa:1994wb,Akmal:1998cf} is employed for nucleonic part of hypernuclear matter while hyperons are considered as non interacting particles. The chemical potential of nucleons has been calculated within LOCV method and the equations of $\beta$-stability in the presence of free hyperons as well as the TOV equations for the mass-radius relation of hypernuclear stars have been solved \cite{Shahrbaf:2019wex}. 
The results for the EoS of nuclear matter (LOCV) and hypernuclear matter (LOCVY) obtained within LOCV method are shown in Fig.~1.

\subsection{Quark matter EoS within nlNJL model}
For the quark matter phase we employ a color superconducting nonlocal chiral quark model of the Nambu--Jona-Lasinio type (nlNJL) for the case of two quark flavors. 
This model has recently been discussed in the context of current compact star constraints in Ref.~\cite{Alvarez-Castillo:2018pve}, where also references to preceding work are given. For the convenience of the reader we summarize here the nlNJL approach which is characterized by four-fermion interactions in the scalar quark-antiquark, the anti-triplet scalar diquark and the vector quark-antiquark channels.
The effective Euclidean action for two light flavors reads
\begin{eqnarray}
S_E &=& \int d^4 x \ \left\{ \bar \psi (x) \left(- i \rlap/\partial + m_c
\right) \psi (x) - \frac{G_S}{2} j^f_S(x) j^f_S(x) 
\right.\nonumber\\
&& \left. 
- \frac{H}{2}
\left[j^a_D(x)\right]^\dagger j^a_D(x) {-}
\frac{G_V}{2} j_V^{\mu}(x)\, j_V^{{\mu}}(x) \right\} \, .
\label{action}
\end{eqnarray}
We considered the current quark mass, $m_c$, to be equal for $u$
and $d$ quarks. The nonlocal currents $j_{S,D,V}(x)$, based on a separable approximation of the effective one gluon exchange (OGE) model of QCD, read
\begin{eqnarray}
j^f_S (x) &=& \int d^4 z \  g(z) \ \bar \psi(x+\frac{z}{2}) \ \Gamma_f\,
\psi(x-\frac{z}{2})\,,
\\
j^a_D (x) &=&  \int d^4 z \ g(z)\ \bar \psi_C(x+\frac{z}{2}) \ \Gamma_D \ \psi(x-\frac{z}{2}) 
\\
j^\mu_V (x) &=& \int d^4 z \ g(z)\ \bar \psi(x+\frac{z}{2})~ i\gamma^\mu
\ \psi(x-\frac{z}{2}), 
\label{cuOGE}
\end{eqnarray}
where $\psi_C(x) = \gamma_2\gamma_4 \,\bar \psi^T(x)$,
$\Gamma_f=(\openone,i\gamma_5\vec\tau)$ and $\Gamma_D=i \gamma_5 \tau_2 \lambda_a$, 
while $\vec \tau$ and $\lambda_a$, with $a=2,5,7$, stand for Pauli and Gell-Mann 
matrices acting on flavor and color spaces, respectively. 
The function $g(z)$ in Eqs.~(\ref{cuOGE}) is a covariant formfactor wich accounts for the nonlocality of the effective quark interactions \cite{GomezDumm:2005hy}. 

The coupling constants ratios $H/G_S$, $G_V/G_S$ are input parameters. 
For OGE interactions in the vacuum, Fierz transformation leads to $H/G_S =0.75$ and $\eta=G_V/G_S = 0.5$. 
However, these value are subject to rather large theoretical uncertainties. 
In fact, thus far there is no strong phenomenological constraint on the ratio $H/G_S$, except for the fact that values
larger than one are quite unlikely to be realized in QCD since they  might lead to color symmetry breaking in the vacuum
\cite{GomezDumm:2005hy}. 
Below, at the end of this subsection, we will consider two schemes of fixing the values of the coupling constants that will be applied in the present work.

The first one, denoted as "model nlNJLA", assumes a set of coupling constants to be fixed for vacuum conditions and afterwards, at finite densities remains unchanged.
The second one is denoted as "model nlNJLB" and has been introduced in Ref.~\cite{Alvarez-Castillo:2018pve} as a generalized nlNJL model with a functional dependence of parameters on the baryochemic potential as the natural thermodynamic variable of the pressure as thermodynamic potential in the grand canonical ensemble.
Before giving further details on models nlNJLA and nlNJLB, we describe the mean field approximation to the thermodynamic potential of the nlNJL model.

After a proper bosonization of this quark model, introducing scalar, vector, and diquark fields, we work in mean field approximation (MFA). 
In the diquark sector, owing to the color symmetry, one can rotate in color space to fix $\Delta_5 = \Delta_7 = 0, \Delta_2 = \Delta$. 
Finally, we consider the Euclidean action at zero temperature and finite baryon chemical potential, where we introduce six different chemical potentials $\mu_{fc}$, depending on the to quark flavors $f=u,d$ and quark colors $c=r,g,b$.

The MFA grand canonical thermodynamic potential per unit volume can be written as
\begin{eqnarray}
\Omega^{\rm MF}  &=&   \frac{ \bar
\sigma^2 }{2 G_S} + \frac{ {\bar \Delta}^2}{2 H} 
- \frac{\bar \omega^2}{2 G_V} \nonumber\\
&&- \frac{1}{2} \int \frac{d^4 p}{(2\pi)^4} \ \ln
\mbox{det} \left[ \ S^{-1}(\bar \sigma ,\bar \Delta, \bar \omega,
\mu_{fc}) \right] \ . \label{mfaqmtp}
\end{eqnarray}

A detailed description of the model and the explicit expression for the thermodynamic potential after calculating the determinant of the inverse of the propagator can be found in Ref.\cite{Blaschke:2007ri}.
The mean field values $\bar \sigma$, $\bar \Delta$ and $\bar \omega$ satisfy the coupled equations
\begin{eqnarray}
\frac{ d \Omega^{\rm MF} }{d\bar \Delta} \ = \ 0 \ , \ \ \
\frac{ d \Omega^{\rm MF} }{d\bar \sigma} \ = \ 0 \ , \ \ \
\frac{ d \Omega^{\rm MF} }{d\bar \omega} \ = \ 0 \ .
\label{gapeq}
\end{eqnarray}

As shown above, there is a freedom in choosing the direction of $\bar\Delta$ in color space. 
In the ansatz we considered, the colors participating in the pairing are $r,g$, leaving the blue color unpaired. 
When considering that our system is in chemical equilibrium, one can see that all chemical potentials are no longer independent, and can be expressed in terms of three independent quantities: the baryonic chemical potential $\mu$, a quark
electric chemical potential $\mu_{Q_q}$ and a color chemical potential $\mu_8$. 
The corresponding relations read
\begin{eqnarray}
\mu_{ur} &=& \mu_{ug} = \frac{\mu}{3} + \frac23 \mu_{Q_q} + \frac13
\mu_8 \ ,\\
\mu_{ub} &=& \frac{\mu}{3} + \frac23 \mu_{Q_q} -
\frac23 \mu_8 \ , \\
\mu_{dr} &=& \mu_{dg} = \frac{\mu}{3} - \frac13 \mu_{Q_q} + \frac13
\mu_8 \ , \\
\mu_{db} &=& \frac{\mu}{3} - \frac13 \mu_{Q_q} - \frac23 \mu_8 \ \ ,
\label{chemical}
\end{eqnarray}
where the chemical potential $\mu_{Q_q}$ distinguishes between up and down quarks, and the color chemical potential 
$\mu_8$ has to be introduced to ensure color neutrality.

In this work we are interested in describing the behaviour of quark matter in the core of NSs, therefore, we shall consider that quark matter is electrically and color neutral and in equilibrium under weak interactions. 
We include electrons and muons as a free relativistic Fermi gas. 
Beta decay reactions read
\begin{equation}
d\to u+l+\bar\nu_l\ ,~~~ { u+l\to d+\nu_l\ ,}
\end{equation}
for $l=e,\mu$. 
We assume that (anti)neutrinos are no longer trapped in the stellar core.
Then, we have the following relation between chemical potentials
\begin{equation}
\label{betaeq}
\mu_{dc} - \mu_{uc} = - \mu_{Q_q} = \mu_l
\end{equation}
for $c=r,g,b$, $\mu_e = \mu_\mu = \mu_l$.
Finally, we impose electrical and color charge neutrality conditions 
\begin{eqnarray}
\label{charge}
\rho_{Q_{tot}} &=& \rho_{Q_q}- \sum_{l=e,\mu}\rho_l \nonumber \\ 
&=&\sum_{c=r,g,b} \left(\frac23 \rho_{uc} - \frac13 \rho_{dc} \right)
- \sum_{l=e,\mu}\rho_l = 0 \ , \\
\rho_8 & = & \frac{1}{\sqrt3} \sum_{f=u,d}
\left(\rho_{fr}+\rho_{fg}-2\rho_{fb} \right) \ = \ 0 \ ,
\label{dens}
\end{eqnarray}
where the expressions for the lepton densities $\rho_l$ and the quark
densities $\rho_{fc}$ can be found in the Appendix of \cite{Blaschke:2007ri}.

Summing up, for each value of $\mu$ one obtains $\bar \Delta$, $\bar \sigma$, $\mu_l$ and $\mu_8$ by self-consistently solving the gap equations (\ref{gapeq}), together with $\beta-$ equilibrium Eq.~(\ref{betaeq}) and (global) electric and color charge neutrality Eqs. (\ref{charge}) and (\ref{dens}) conditions.
The quark matter EoS is then
\begin{equation}
\label{eq:P-mu}
P(\mu)=P(\mu;\eta(\mu),B(\mu))=-\Omega^{\rm MF}(\eta(\mu)) - B(\mu)~,
\end{equation}
where for later use we allow for the possibility of a bag pressure shift $B$ stemming, e.g., from a medium dependence of the gluon sector, and both parameters $\eta$ and $B$ may depend on the chemical potential \cite{Alvarez-Castillo:2018pve}. In this work we discuss two cases of this generalized nlNJL model for quark matter denoted as model nlNJLA and model nlNJLB.

\begin{figure*}
\centering
\includegraphics[width=\textwidth]{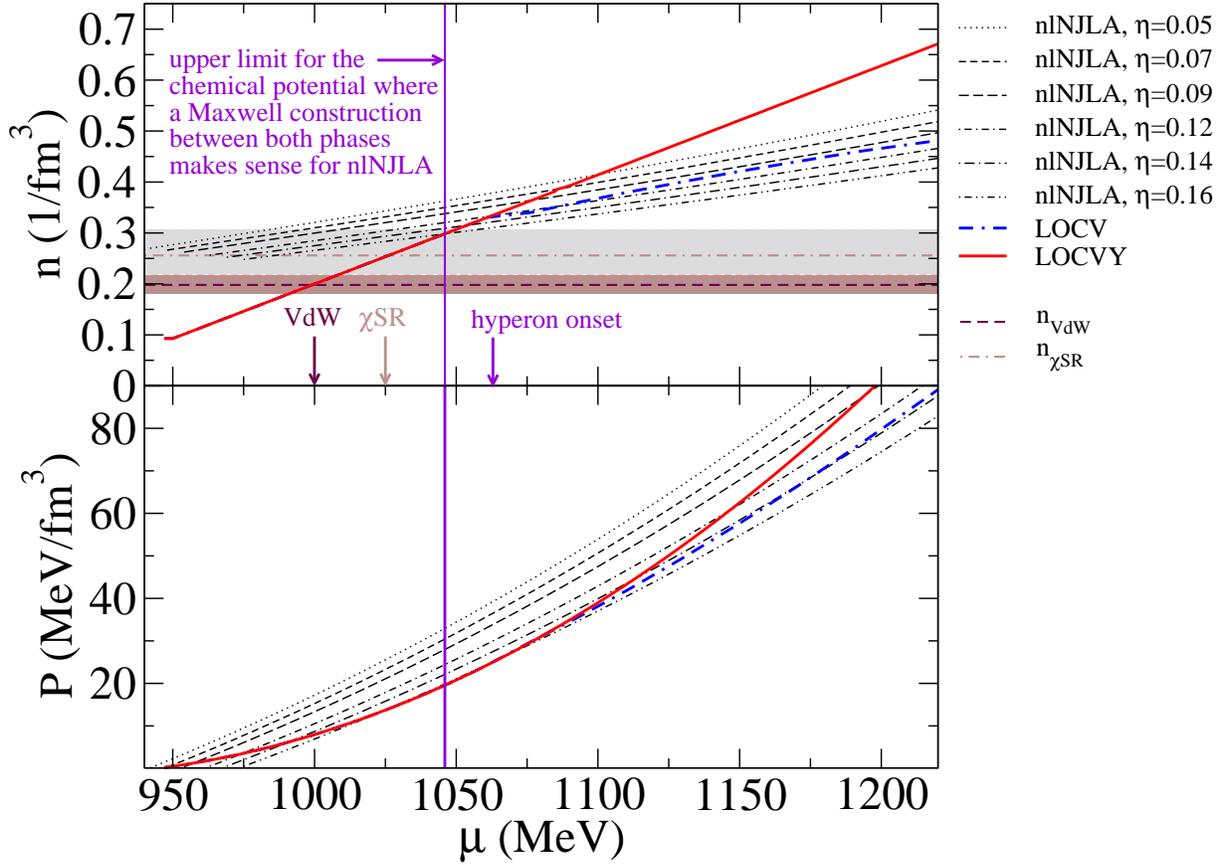}
\caption{\label{fig:1} Nuclear and hypernuclear matter EoS obtained from the LOCV method, compared with quark matter EoS according to the nlNJLA model with color superconductivity for different constant values of the dimensionless vector meson coupling strength parameter $\eta$.
Lower panel: pressure $P$ vs. baryon chemical potential $\mu$; upper panel: baryon density $n$ vs. baryon chemical potential $\mu$. 
The maroon and grey vertical arrows show estimates for the chemical potentials and the corresponding densities 
$n_{VdW}$ and $n_{\chi SR}$ where limits of the applicability of the LOCV(Y) EoS are reached due to the finite size of the nucleons with their Van der Waals volume ("VdW") and the partial chiral symmetry restoration ($\chi SR$), respectively. 
For details, see text.
The violet vertical line indicates the limit of applicability of a Maxwell construction between (hyper)nuclear and quark matter in this case since the density of the baryonic phase as a system of confined quark bound states should always be lower than that of the deconfined quark matter at a critical chemical potential where the pressure curves of both phases cross each other. 
The violet vertical arrow "hyperon onset" indicates the chemical potential for which the LOCVY EoS starts deviating from the LOCV one. }
\end{figure*}

Model nlNJLA is the case when $B(\mu)={\rm const}$ and $\eta(\mu)= {\rm const}$. 
In Fig.~\ref{fig:1} (lower panel) we show the resulting EoS for color superconducting quark matter for different values of $\eta$ in comparison with the ones for nuclear and hypernuclear matter. 
According to the maximum entropy principle of equilibrium thermodynamics in the presence of an alternative  phase of matter under given conditions, the one with the higher pressure shall be realised.  
At the crossing of the corresponding lines for $P(\mu)$ the Gibbs conditions for phase equilibrium (with global charge conservation) are fulfilled and one can obtain the Maxwell construction for an equilibrium phase transition 
(see the next subsection) by following the curve with the higher pressure.

However, by inspecting the lower panel of Fig.~\ref{fig:1} we realise that the curves for the quark matter pressure cross the ones for (hyper)nuclear matter twice. While the first crossing at lower chemical potentials corresponds to a hadron-to-quark matter PT, the second one would correspond to a transition from quark to hadronic matter (reconfinement) when increasing 
the density. 
Such a transition should be excluded on the basis of the understanding that the hadronic phase with baryons as elementary degrees of freedom in the theory of strongly interacting matter (like in the LOCV(Y) approach) is limited to low densities where the quark degrees of freedom are confined and thus localized inside the hadrons as their bound states.
By increasing the density a transition to the deconfined phase of quark matter takes place that physically corresponds to a delocalization of the quark wave functions and the transition to a basis of quasifree quark quasiparticle states. 
The physical reason for this transition is the necessity to fulfill the Pauli principle on the quark level of description when the nucleon wave functions start overlapping. Increasing the density further the consequences of the necessity to antisymmetrize the quark wave functions will be even more severe and will lead to the formation of a conduction band of delocalized states of the many-quark wave function. 
The reconfinement transition is excluded because once the transition to deconfined quark matter with the corresponding 
change of the physical basis states has been made, the LOCV(Y) EoS is no longer applicable since it is based on the use of many-nucleon wave functions with quarks being confined in the nucleons.

In the upper panel of Fig.~\ref{fig:1} we show the densities that correspond to the pressures in the lower panel and 
are obtained from them as derivative $n(\mu)=dP(\mu)/d\mu$. 
The density is a suitable quantity to estimate the limits of the validity of hadronic EoS like the LOCV(Y) approach.
We discuss two examples for such limitations.
The first limitation is due to the finite size of nucleons and can be quantified by the maroon dashed line that shows the density 
$n_{VdW}=1/V_{VdW}$, for which on the average one nucleon can be found within the Van der Waals excluded volume 
$V_{VdW}=16\pi r_N^3/3$. The latter arises for a gas of nucleons as hard spheres with a radius that is estimated by the strong interaction radius of nucleons $r_N=0.67 \pm 0.2$ fm \cite{Povh:1990ad}. 
The $1\sigma$ error for $r_N$ translates to the brown uncertainty band in densities 
$n_{VdW}=0.20 \pm 0.018$ fm$^{-3}$ 
and the corresponding region in chemical potentials around the central value $\mu=1.0$ GeV indicated by the maroon 
vertical arrow that is obtained from the crossing points with the red solid line of the LOCV(Y) EoS.  

The second limitation stems from the fact that at finite density the chiral condensate melts which 
according to the Brown-Rho scaling \cite{Brown:1991kk} amounts to a factor 1/2 already at the saturation density.
Here we quantify this limitation by applying the chiral perturbation theory formula for the chiral condensate 
\begin{equation}
\langle \bar q q \rangle = \langle \bar q q \rangle_{\rm vac} \left[1 - \frac{\sigma_{\pi N} n_{s,N}}{f_\pi^2 M_\pi^2} \right], 
\end{equation}
with $M_\pi = 140$ MeV and $f_\pi=0.093$ MeV being the vacuum pion mass and decay constant, respectively.
Because of the relatively low densities, we may identify the  scalar nucleon density $n_{s,N}$ with the baryon density $n$
and use the N$^3$LO result from chiral perturbation theory for the pion-nucleon sigma term $\sigma_{\pi N}=43\pm 7$ MeV 
(see, e.g., Ref.~\cite{Jankowski:2012ms} for details)
to obtain a 50\% reduction of the chiral condensate at the density $n_{\chi SR} = 0.256^{+0.050}_{-0.036}$ fm$^{-3}$.
The mean density value for this limit due to partial chiral symmetry restoration corresponds to a chemical potential of
$1025$ MeV indicated by a brown vertical arrow in the upper panel of Fig.~\ref{fig:1}. 

Chiral symmetry restoration in the hadronic phase of matter leads to a parity doubling of hadronic states. For the applications to compact stars this concerns in particular the chiral partner state of the nucleon, the N(1535) which at a density exceeding $n_{\chi SR}$ will become degenerate in mass with the nucleon. 
This scenario, supported by lattice QCD simulations at finite temperatures \cite{Aarts:2017rrl},  
has been investigated, e.g., in Ref.~\cite{Marczenko:2018jui} and can lead to a first order phase transition to chirally symmetric nuclear matter. 
Thus chiral symmetry restoration invalidates the application of the LOCV(Y) which does not account for such a medium-dependence of the hadronic basis states.

We would like to summarize the discussion of the reconfinement problem. 
Following the examples discussed above for physical limitations of the LOCV(Y) EoS, we should abstain from applying it when the chemical potential exceeds about 1050 MeV. 
However, as our estimates are rather crude and in the absence of a better alternative (such as a first principle calculation based on QCD at high-density) we may have to use the LOCV(Y) EoS even beyond that point, in particular for a discussion of the onset of hyperons in this EoS which lies at  $\mu_{\rm onset}=1063$ MeV and is indicated by the violet vertical arrow
in the upper panel of Fig.~\ref{fig:1}.
Then another, more formal condition for excluding reconfinement may be applied which is based on the Gibbs conditions for the Maxwell construction of the phase transition (see the next subsection).
One should ignore those crossing points of the quark matter $P(\mu)$ curves with the hadronic EoS at which the slope (corresponding to the density) of the hadronic curve exceeds that of the quark matter one since they would correspond to a quark-to-hadron matter transition (reconfinement).
The limiting chemical potential for which the hadronic EoS can be applied for a reasonable Maxwell construction of the  phase transition is then the one at which the slopes of the hadronic and quark matter $P(\mu)$ curves coincide. 
This condition for the applicability of the hadronic EoS depends now also on the choice of the quark matter EoS and it is indicated for model nlNJLA by a violet vertical line in Fig.~\ref{fig:1}.

Note that for model nlNJLA  the deconfinement is at rather low chemical potentials, even lower than our estimates for the limitations of the validity of the LOCV(Y) EoS. 
We attribute this to the lack of confining effects in this EoS. 
In model nlNJLB we will add a confining effect in the thermodynamics by a bag pressure $B(\mu)$ that parametrically depends on the chemical potential, i.e. the density of the medium. 

The model nlNJLB is the general case with a density-dependence for both, the vector meson coupling strength 
$\eta(\mu)$ and confining bag function $B(\mu)$.
The density dependence of these two parameters of the generalized nlNJL model has been introduced 
in Ref.~\cite{Alvarez-Castillo:2018pve} in the form
\begin{equation}
\label{eq:eta-mu}
\eta(\mu)=\eta_> f_\gg(\mu)+ \eta_< f_\ll(\mu)
\end{equation} 
and
\begin{equation}
\label{eq:bag}
B(\mu)=Bf_<(\mu)f_\ll(\mu)~,
\end{equation} 
which follows when the pressure is constructed by interpolating\footnote{We note that this interpolation between different parametrizations of the nlNJL model for quark matter shall not be confused with the interpolation between quark and hadronic asymptotic EoS as described, e.g., in \cite{Masuda:2012ed}
following the intentions of \cite{Asakawa:1995zu}.
Here, we employ the interpolation technique of Ref.~\cite{Blaschke:2013rma} which has also been used in 
\cite{Alvarez-Castillo:2018pve} and denote the resulting description of the quark matter EoS as model nlNJLB. }
between three nlNJL pressures with constant coefficients, 
\begin{eqnarray}
\label{eq:twofold2}
P(\mu) &=& [f_<(\mu) P(\mu;\eta_<,B) + f_>(\mu) P(\mu;\eta_<,0)] f_{\ll}(\mu)
\nonumber\\
&& +f_{\gg}(\mu)P(\mu;\eta_>,0)~,
\end{eqnarray}
with the switching functions
\begin{equation}
\label{eq:f<}
f_{<}(\mu)=\frac{1}{2}\left[1-\tanh\left(\frac{\mu-\mu_<}{\Gamma_<}\right)\right]~,\\
f_{>}(\mu) = 1 - f_<(\mu)~,
\end{equation}
and $f_{\ll}(\mu)$, $f_{\gg}(\mu)$ being defined analogous to (\ref{eq:f<}) by replacing $\mu_< \to \mu_\ll$
and $\Gamma_< \to \Gamma_\ll$.
In the present work we will employ four sets of parameters for constructing the density dependence of the coefficients
in model nlNJLB that are given below in table \ref{tab:tableI}.  
\begin{table}[h]
 \caption{\label{tab:tableI}
 Four different parameter sets which define the density-dependent coefficients of the model nlNJLB.}
 \begin{ruledtabular}
 \begin{tabular}{lccccc}
 \textrm{parameter}&\textrm{set 1}&\textrm{set 2}&\textrm{set 3}&\textrm{set 4}
 \\
 \\
 \hline
 {$\mu_< $ (MeV)} & 1090 & 1090 & 1090 & 1070  \\
{$\Gamma_<$ (MeV)} & 155 & 163 & 150 & 170 \\
{$\mu_\ll$ (MeV)} & 1500 & 1500 & 1500 & 1600 \\
{$\Gamma_\ll $ (MeV)} & 300 & 300 & 270 & 300 \\
{$\eta_<$} & 0.05 & 0.05 &0.07 & 0.07 \\
{$\eta_>$} & 0.09 & 0.12 & 0.12 & 0.16 \\
{B ({MeV}/{fm$^3$})} & 30 & 30 & 20 & 25 \\
 \end{tabular}
 \end{ruledtabular}
 \end{table}
 
For high densities, above a matching point that in the present work is fixed to be at 
$\varepsilon_{\rm CSS}=690$ MeV/fm$^3$, we replace the EoS of model nlNJLB with a constant speed of sound (CSS) 
model as it has been introduced, e.g., in Refs.~\cite{Alford:2013aca,Zdunik:2012dj} and was used extensively in recent works on the classification of hybrid compact stars (see \cite{Alford:2015gna} and references therein) and their appearance in third \cite{Paschalidis:2017qmb} and fourth \cite{Alford:2017qgh} families of stability branches of compact stars. 
In \cite{Zdunik:2012dj} it was demonstrated that NJL model based approaches to color-superconducting, cold quark matter 
can be very well approximated by a CSS parametrization.
The matching is performed such that the pressure and pressure derivative of the CSS model and the nlNJLB model coincide
at the matching point the position of which is a matter of choice. 
The CSS extrapolation became necessary because of a limitation of the covariant formfactor realization of the nlNJL model 
\cite{GomezDumm:2005hy} to not too high densities. 

\subsection{Phase transition construction}
For constructing a first order PT between hadronic phase and deconfined quark matter we will apply here the Maxwell construction (MC) which assumes that both EoS should separately fulfill the charge-neutrality and $\beta$-equilibrium conditions with electrons and muons. 
Under such conditions the chemical potential of a particle species $i$ can be written as
\begin{equation}
\mu_i=b_i\mu+q_i\mu_q
 \end{equation}
where $b_i$ is the baryon number of the species $i$, $q_i$ denotes its charge in units of the electron charge, $\mu$ and $\mu_q$ are the baryonic and electric chemical potentials, respectively.
The  Gibbs conditions for phase equilibrium require that the temperatures, chemical potentials and pressures of the two phases should coincide at the phase transition                                    
\begin{eqnarray}
\mu_H&=&\mu_Q=\mu_c~,\\
T_H&=&T_Q=T_c~,\\
P_H(\mu_B,\mu_e)&=&P_Q(\mu_B,\mu_e)=P_c~.
\end{eqnarray}
In above equations, the subscript $c$ denotes the critical value of these thermodynamic variables for which chemical, thermal and mechanical phase equilibrium is established. 
In the present work we construct the PT for the zero temperature case. 
Technically, one can plot the pressure as a function of chemical potential for two phases and merge them at the crossing point in which $P_{\mu_c}=P_c$. 
Physically, the phase with higher pressure (lower grand canonical potential) in a given chemical potential, is considered out of PT region. 
We note that this setting of the problem inevitably evokes the so-called "reconfinement problem"  
\cite{Lastowiecki:2011hh,Zdunik:2012dj}. 
In the present work, we will employ the "no reconfinement" paradigm which states that once the critical chemical potential for the deconfinement transition has been reached the "old" hadronic EoS will no longer be considered as a relevant alternative 
at still higher chemical potentials.
Therefore, should a second crossing between hadronic and quark matter EoS occur in the pressure vs. chemical potential plane it shall be ignored and thus a "reconfinement" will be excluded.

\section{Mass-radius relation for hybrid stars}  
A particularly interesting question in this context is the role which hyperons or deconfined quark matter can play in interpreting the observations of binary neutron star mergers. 
The observational constraints for NS maximum mass and radius from the millisecond pulsar PSR J0740+6620 and the binary neutron star merger GW170817 should be fulfilled by the theoretical calculations.
We consider the hybrid star as a hydrostatically equilibrated and spherically symmetric system. 
Thus, the mass-radius relation of the star can be obtained for a given EoS by solving the well-known Tolman-Oppenheimer-Volkoff (TOV) equations  \cite{Tolman:1939jz,Oppenheimer:1939ne}, 
\begin{eqnarray}
\frac{dP(r)}{dr} &=&-\frac{GM(r)\varepsilon (r)}{c^{2} r^{2} } \left(1+\frac{P(r)}{\varepsilon (r)} \right) 
\left(1+\frac{4\pi r^{3} P(r)}{M(r)c^{2} } \right)
 \left(1-\frac{2GM(r)}{rc^{2} } \right)^{-1},\\
\frac{dM(r)}{dr} &=&\frac{4\pi \varepsilon (r)r^{2} }{c^{2} }.
\end{eqnarray}
In these equations $P(r)$ and $\varepsilon(r)$ denote the pressure and the energy density profiles for the matter distribution in the CS interior, $M(r)$ is the cumulative mass enclosed in a spherical volume at the distance $r$ from the center, and $G$ is the gravitational constant. The gravitational mass  $M=M(r=R)$  of the star is the mass enclosed within the radius of the star. 
By considering that the boundary condition $P(r=R)=0$ defines the radius $R$ for a chosen central energy density 
$\varepsilon_c=\varepsilon(r=0)$, we have the necessary boundary and initial conditions to solve the TOV equations for a relativistic star with mass $M$ and radius $R$, respectively. 
By increasing $\varepsilon$ (or equivalently $P$) up to the maximum mass, the mass-radius relation can be obtained. 
For the EoS of the inner and outer crust of the neutron star, we use the results of Negele and Vautherin \cite{Negele:1971vb} and Harrison and Wheeler \cite{harrisonwheeler}, respectively.

\section{Results}

Fig. 2 and Fig. 3 show the EoS for pressure as a function of chemical potential and energy density respectively for the Maxwell construction of the deconfinement phase transition. 
As it can be seen in the Fig. 3, the jump in energy density increases by decreasing the vector meson coupling parameter. 
However, the physically relevant crossing points occur at a low chemical potentials, before the onset of hyperons
at $\mu=1063$ MeV.

The gravitational mass of the hybrid star in the unit of solar mass has been plotted in Fig. 4 as a function of radius. 
As it was mentioned before, we have considered the PT to quark matter as a solution for hyperon puzzle. 
The figure shows that the PT to deconfined quark matter which we have constructed not only increases the maximum mass 
of the hybrid star branch when compared to that of the hyperonic star sequence, but also all hybrid star EoS fulfill the new observational constraint for the maximum mass of CS from PSR J0740+6620 \cite{Cromartie:2019kug}.

Fig. 5 shows an example how the generalized nonlocal chiral quark model EoS (nlNJLB) with $\mu$-dependent bag function $B(\mu)$ and vector coupling $\eta(\mu)$  is obtained for the set 2 of Table \ref{tab:tableI}, using three parametrizations of the model nlNJLA with density-independent coefficients is an input.

It is worth mentioning that for very high density, we have applied the constant speed of sound (CSS) extrapolation method to reach the maximum mass of hybrid stars. This technique can be considered as the lowest-order terms of a Taylor expansion of the quark matter EoS about the transition pressure \cite{Alford:2013aca}. An example of this extrapolation method for the set 2 of Table \ref{tab:tableI} is shown in Fig. 6. As it can be seen in the figure, there is a good agreement between the nlNJLB EoS and the extrapolated one at high densities. The position of the matching point for this parametrization is also shown.

Fig. 7 shows the squared speed of sound $c_s^2=dP/d\varepsilon$ as a function of the energy density for all sets in table~\ref{tab:tableI} that the CSS extrapolation method has been used for them in high chemical potential to reach the maximum mass of hybrid star. 
The figure shows the regions of the first-order PT where $c_s^2$ in unit of the speed of light squared is equal to zero and fulfils the condition of causality $c_s^2<1$ at all energy densities. One should note that the quark matter speed of sound is held constant after matching point at $\varepsilon=690$ MeV/fm$^3$ which for nlNJLB set 2 corresponds to $\mu=1362$ MeV.

Fig. 8 and Fig. 9 show the pressure as a function of chemical potential and energy density, respectively, for the Maxwell construction of the deconfinement PT in which the LOCVY EoS for hypernuclear matter as well as the four sets of interpolated (with extrapolation at high chemical potential) EoS for quark matter have been used. 
The gravitational mass of the hybrid star in the unit of solar mass has been plotted in Fig. 10 as a function of radius. 
As shown in the figures 8-10, by using the interpolation method of model nlNJLB, we obtain a strong PT for which the onset of deconfinement takes place at  a sufficiently large chemical potential so that there is an intermediate hypernuclear phase between the nuclear and the deconfined quark matter phases. 

Furthermore, the obtained maximum masses obey the current CS constraints:
$M_{\rm max} > 2.07~M_\odot$ from the lower limit of the $1\sigma$ range of the Shapiro-delay-based mass measurement for the millisecond pulsar PSR J0740+6620 \cite{Cromartie:2019kug}, as well as the lower and upper limits for the radius, $R_{1.6M_\odot}>10.7$ km \cite{Bauswein:2017vtn} 
and $R_{1.4M_\odot}<13.6$ km \cite{Annala:2017llu}, respectively,  from the binary CS merger GW170817 \cite{TheLIGOScientific:2017qsa}.
It should be mentioned that for all cases considered in the present work, the hypernuclear EoS is not stiff enough at low densities so as to produce a sufficiently large jump in the energy density tat the deconfinement transition for obtaining a disconnected third family of CSs in the M-R diagram. 
For details on the relation of the classification of hybrid star sequences and their relation to the size of the jump in energy density and the critical pressure at the phase transition, see Ref.~\cite{Alford:2013aca}.

In Fig. 11 we show the profiles of energy densities for the model nlNJLA (upper panel) with four cases of vector coupling strength $\eta=0.12, 0.13, 0.14, 0.15$ and the four sets of parametrisation of model nlNJLB (lower panel) for the case of a typical binary radio pulsar mass of $1.35~M_\odot$ (which is relevant for scenarios explaining the binary merger GW170817).
While for all parametrizations of model nlNJLA the deconfinement transition occurs directly from the nuclear matter outer core to the extended quark matter core, in model nlNJLB there is a shell of hypernuclear matter in-between the inner core of superconducting  quark matter and the outer core of nuclear matter.

In Fig. 12 we show the same comparison of compact star profiles as in Fig. 11, just for a pulsar mass of $2.0~M_\odot$.
In comparison to the case of pulsar mass of $1.35~M_\odot$ shown in Fig. 11 now the core of color superconducting quark matter is much larger.

Finally, in Fig. 13 we discuss the case of isospin symmetric matter that is relevant for applications to relativistic heavy-ion collisions. 
In this case, due to the short lifetime of the fireball created under these conditions, there is no $\beta-$ equilibrium constraint with respect to weak interactions and effects of Coulomb interactions can be neglected when compared to those of strong interactions. 
We see that for the model nlNJLA (upper panel) the cases with sufficiently repulsive vector meson mean field ($\eta > 0.12 $)
do predict a deconfinement transition at high densities beyond $0.79$ fm$^{-3}$. 
The transitions from quark to hadronic  phase at lower densities are unphysical and shall be ignored.
Their appearance may be attributed to the absence of a confining mechanism for quarks in model nlNJLA. 
Less repulsive vector mean fields ($\eta \le 0.09 $) do not predict a deconfinement transition by the same reason. 
For the model nlNJLB (shown in the lower panel of Fig. 13) all considered parametrizations (set 1 - set 4) predict a deconfinement phase transition under isospin-symmetric conditions with onset densities between 2.2 and 2.7 $n_0$. 
Under compact star conditions these parametrizations of model nlNJLB predict onset masses for quark 
deconfinement between $0.99~M_\odot$ and $1.14~M_\odot$ while fulfilling the maximum mass constraint and thus 
solving the hyperon puzzle. 
As a characterizing feature of this hybrid EoS model, there is an intermediate hypernuclear phase for all stars in the 
range of observed compact star masses. 

\section{Conclusion} 
We have reconsidered the problem that the appearance of hyperons softens the nuclear EoS such that under compact star 
conditions the constraint on the lower limit for the maximum mass at $2~M_\odot$ can not be fulfilled (hyperon puzzle).
To this end we have applied a two-phase approach to hybrid compact star matter where the hadronic phase is described using the LOCV method with AV18 and Urbana three-body interaction including hyperons and the quark matter phase is modelled by a color superconducting nonlocal Nambu-Jona-Lasinio model with constant (model nlNJLA) and with density-dependent (model nlNJLB) parameters. 
The phase transition has been obtained by a Maxwell construction. 
Our study confirms that also with the present set of equations of state the quark deconfinement presents a viable 
solution of the hyperon puzzle. A new finding of the present work is that the model nlNJLB allows for an intermediate hypernuclear matter phase in the hybrid star, between the nuclear and color superconducting quark matter phase, while in the model nlNJLA such a phase cannot be realized because the phase transition onset is at low densities, before the hyperon threshold density is passed. 

We have discussed the possible application of the present hybrid EoS for estimating the onset density of the deconfinement phase transition in symmetric matter as it will be probed in future heavy-ion collision experiments at FAIR, NICA and corresponding energy scan programs at the CERN and RHIC facilities. 
We found that for the model nlNJLA the cases with a sufficiently strong repulsive vector mean field ($\eta > 0.12 $) which produce reasonable hybrid star EoS have also a phase transition under isospin-symmetric conditions. 
For $\eta = 0.12$ the critical density is $n_c=0.79$ fm$^{-3}$ and for $\eta = 0.15$ it is $n_c=0.98$ fm$^{-3}$.
For the less repulsive vector mean fields ($\eta \le 0.11 $) there is no deconfinement transition in symmetric matter!
%
%
This may be attributed to the absence of a confining mechanism for quarks in the model nlNJLA. 
For the model nlNJLB a density-dependent bag pressure serves as a confining mechanism at low densities.
This model predicts a deconfinement phase transition under isospin-symmetric conditions for all considered 
parametrizations (set 1 - set 4) at densities between 2.2 $n_0$ and 2.7  $n_0$.  
For these same parameter sets the hybrid star EoS
predicts an onset of quark deconfinement for compact stars in the mass range from $0.99~M_\odot$ to $1.14~M_\odot$ 
while fulfilling the maximum mass constraint and thus solving the hyperon puzzle. 
It is remarkable that in this model all compact stars in the observed range of masses (i.e. from about 1.2 to 2.2 $M_\odot$) 
may have a shell of hypernuclear matter between the inner core of color superconducting quark matter and the outer core of nonstrange nuclear matter.
Therefore, the scenarios of binary compact star mergers should consider the case of coalescing hybrid stars and investigate 
the role of hypernuclear and quark matter phases in this context.

The application of the quark-hadron hybrid EoS to binary merger simulations, however, requires the inclusion of finite temperatures which is planned for the future extension of the presented approach. 
On such a basis the recently suggested gravitational wave signal for a deconfinement transition \cite{Bauswein:2018bma}
could then be discussed and the supernova explodability of blue supergiant stars \cite{Fischer:2017lag}
when described with the finite-temperature extension of the here introduced class of hybrid EoS could be investigated.
To such a further development of the EoS would correspond a structure of the QCD phase diagram in the full space of variables, i.e., temperature, baryon density, and isospin asymmetry 
which besides astrophysical applications is of relevance for simulations of  heavy-ion collision experiments. 

\subsection*{Acknowledgements}
We acknowledge discussions with D. Alvarez-Castillo, A. Ayriyan, A. Bauswein, T. Fischer, K. Maslov, A. Sedrakian, 
S. Typel and D.N. Voskresensky and their helpful comments on the first version of the manuscript.
M.S. is grateful to the research council of the University of Tehran for supporting her research visit at the University of 
Wroclaw where this work has been performed, and to the HISS Dubna program for supporting her participation in the 
summer schools in 2018 and 2019 where this project was initiated and completed, also with the support of the 
Bogoliubov-Infeld program for scientist exchange between Polish Institutes and JINR Dubna. 
The work of D.B. was supported by the Russian Science Foundation under grant number 17-12-01427.
D.B and M.S. thank the European COST Action CA16214 "THOR" for supporting their networking activities.


\begin{thebibliography}{1}

         \bibitem{Steiner:2010fz}
             A.~W.~Steiner, J.~M.~Lattimer and E.~F.~Brown,
             Astrophys.\ J.\  {\bf 722} (2010) 33.

\bibitem{Cromartie:2019kug}
  H.~T.~Cromartie {\it et al.},
  Nature Astron. 3 (2019) 439; arXiv:1904.06759.
  
\bibitem{TheLIGOScientific:2017qsa}
  B.~P.~Abbott {\it et al.} [LIGO Scientific and Virgo Collaborations],
  Phys.\ Rev.\ Lett.\  {\bf 119} (2017) no.16,  161101.
  
\bibitem{Bauswein:2017vtn}
  A.~Bauswein, O.~Just, H.~T.~Janka and N.~Stergioulas,
  Astrophys.\ J.\  {\bf 850} (2017) no.2,  L34.
  
  \bibitem{Annala:2017llu}
    E.~Annala, T.~Gorda, A.~Kurkela and A.~Vuorinen,
    Phys.\ Rev.\ Lett.\  {\bf 120} (2018) no.17,  172703.
\bibitem{Ambartsumyan:1960}
   V. A. Ambartsumyan and G. S. Saakyan,
   Soviet Astronomy {\bf 4} (1960) no. 2, 187. 
\bibitem{Baldo:1998hd}
   M.~Baldo, G.~F.~Burgio and H.~J.~Schulze,
   Phys.\ Rev.\ C {\bf 58} (1998) 3688.
\bibitem{Baldo:1999rq}
  M.~Baldo, G.~F.~Burgio and H.~J.~Schulze,
  Phys.\ Rev.\ C {\bf 61} (2000) 055801.
  \bibitem{Shahrbaf:2019wex}
      M.~Shahrbaf and H.~R.~Moshfegh,
      Annals Phys.\  {\bf 402} (2019) 66.
   
\bibitem{Yamamoto:2015lwa}
  Y.~Yamamoto, T.~Furumoto, N.~Yasutake and T.~A.~Rijken,
  Eur.\ Phys.\ J.\ A {\bf 52} (2016) no.2,  19
  
\bibitem{vanDalen:2014mqa}
  E.~N.~E.~van Dalen, G.~Colucci and A.~Sedrakian,
  Phys.\ Lett.\ B {\bf 734} (2014) 383.

  
\bibitem{Maslov:2015msa}
  K.~A.~Maslov, E.~E.~Kolomeitsev and D.~N.~Voskresensky,
  Phys.\ Lett.\ B {\bf 748} (2015) 369.
  
\bibitem{Maslov:2015wba}
  K.~A.~Maslov, E.~E.~Kolomeitsev and D.~N.~Voskresensky,
  Nucl.\ Phys.\ A {\bf 950} (2016) 64.
  
\bibitem{Li:2018qaw}
  J.~J.~Li, A.~Sedrakian and F.~Weber,
  Phys.\ Lett.\ B {\bf 783} (2018) 234.
  
  
\bibitem{Providencia:2018ywl}
    C.~Provid\'encia, M.~Fortin, H.~Pais and A.~Rabhi,
    Front. Astron. Space Sci. {\bf6} (2019) 13.

\bibitem{Drago:2014oja}
  A.~Drago, A.~Lavagno, G.~Pagliara and D.~Pigato,
  Phys.\ Rev.\ C {\bf 90} (2014) no.6,  065809.

\bibitem{Vidana:2017qey}
  I.~Vida\~{n}a, M.~Bashkanov, D.~P.~Watts and A.~Pastore,
  Phys.\ Lett.\ B {\bf 781} (2018) 112.

\bibitem{Mukherjee:2017jzi}
  A.~Mukherjee, S.~Schramm, J.~Steinheimer and V.~Dexheimer,
  Astron.\ Astrophys.\  {\bf 608} (2017) A110.
  
\bibitem{Marczenko:2018jui}
  M.~Marczenko, D.~Blaschke, K.~Redlich and C.~Sasaki,
  Phys.\ Rev.\ D {\bf 98} (2018) no.10,  103021.

\bibitem{Marczenko:2019trv}
  M.~Marczenko, D.~Blaschke, K.~Redlich and C.~Sasaki,
  Universe {\bf 5} (2019) 180.


\bibitem{Weber:2004kj}
  F.~Weber,
  Prog.\ Part.\ Nucl.\ Phys.\  {\bf 54} (2005) 193.

\bibitem{Burgio:2002sn} 
  G.~F.~Burgio, M.~Baldo, P.~K.~Sahu and H.~J.~Schulze,
  Phys.\ Rev.\ C {\bf 66} (2002) 025802.

\bibitem{Baldo:2003vx}
  M.~Baldo, G.~F.~Burgio and H.-J.~Schulze,
  astro-ph/0312446. 
  
 
\bibitem{Klahn:2006iw}
         T.~Kl\"ahn, D.~Blaschke, F.~Sandin, C.~Fuchs, A.~Faessler, H.~Grigorian, G.~R\"opke and J.~Tr\"umper,
         Phys.\ Lett.\ B {\bf 654} (2007) 170.
         
\bibitem{Blaschke:2005uj}
  D.~Blaschke, S.~Fredriksson, H.~Grigorian, A.~M.~Oztas and F.~Sandin,
  Phys.\ Rev.\ D {\bf 72} (2005) 065020.
   
\bibitem{Typel:2009sy}
  S.~Typel, G.~R\"opke, T.~Kl\"ahn, D.~Blaschke and H.~H.~Wolter,
  Phys.\ Rev.\ C {\bf 81} (2010) 015803.

\bibitem{Lastowiecki:2011hh}
  R.~Lastowiecki, D.~Blaschke, H.~Grigorian and S.~Typel,
  Acta Phys.\ Polon.\ Supp.\  {\bf 5} (2012) 535.
   
\bibitem{Bonanno:2011ch} 
  L.~Bonanno and A.~Sedrakian,
  Astron.\ Astrophys.\  {\bf 539} (2012) A16.
  
\bibitem{Lalazissis:1996rd}
  G.~A.~Lalazissis, J.~K\"onig and P.~Ring,
  Phys.\ Rev.\ C {\bf 55} (1997) 540.

\bibitem{Lalazissis:2005de}
  G.~A.~Lalazissis, T.~Niksic, D.~Vretenar and P.~Ring,
  Phys.\ Rev.\ C {\bf 71} (2005) 024312.
  
\bibitem{Colucci:2013pya}
  G.~Colucci and A.~Sedrakian,
  Phys.\ Rev.\ C {\bf 87} (2013) 055806.

\bibitem{Shahrbaf:2019bef} 
  M.~Shahrbaf, H.~R.~Moshfegh and M.~Modarres,
  Phys.\ Rev.\ C {\bf 100}, no. 4, 044314 (2019).
   
\bibitem{Alvarez-Castillo:2018pve}
  D.~E.~Alvarez-Castillo, D.~B.~Blaschke, A.~G.~Grunfeld and V.~P.~Pagura,
  Phys.\ Rev.\ D {\bf 99} (2019) no.6,  063010.

\bibitem{Povh:1990ad} 
  B.~Povh and J.~H\"ufner,
  Phys.\ Lett.\ B {\bf 245}, 653 (1990).
  
\bibitem{Brown:1991kk} 
  G.~E.~Brown and M.~Rho,
  Phys.\ Rev.\ Lett.\  {\bf 66}, 2720 (1991).

\bibitem{Jankowski:2012ms} 
  J.~Jankowski, D.~Blaschke and M.~Spalinski,
  Phys.\ Rev.\ D {\bf 87}, no. 10, 105018 (2013).

\bibitem{Aarts:2017rrl} 
  G.~Aarts, C.~Allton, D.~De Boni, S.~Hands, B.~J\"ager, C.~Praki and J.~I.~Skullerud,
  JHEP {\bf 1706}, 034 (2017).


\bibitem{Marczenko:2018jui} 
  M.~Marczenko, D.~Blaschke, K.~Redlich and C.~Sasaki,
  Phys.\ Rev.\ D {\bf 98}, no. 10, 103021 (2018).


\bibitem{Ayriyan:2018blj}
  A.~Ayriyan, D.~Alvarez-Castillo, D.~Blaschke and H.~Grigorian,
  Universe {\bf 5} (2019) no.2,  61.
    
  \bibitem{GomezDumm:2005hy} 
  D.~Gomez Dumm, D.~B.~Blaschke, A.~G.~Grunfeld and N.~N.~Scoccola,
  Phys.\ Rev.\ D {\bf 73} (2006) 114019.
  
\bibitem{Blaschke:2007ri} 
  D.~B.~Blaschke, D.~Gomez Dumm, A.~G.~Grunfeld, T.~Kl\"ahn and N.~N.~Scoccola,
  Phys.\ Rev.\ C {\bf 75} (2007) 065804.
  
\bibitem{Kaltenborn:2017hus}
  M.~A.~R.~Kaltenborn, N.~U.~F.~Bastian and D.~B.~Blaschke,
  Phys.\ Rev.\ D {\bf 96} (2017) no.5,  056024.
  
\bibitem{Gerlach:1968zz}
  U.~H.~Gerlach,
  Phys.\ Rev.\  {\bf 172} (1968) 1325.
  
  \bibitem{Alford:2013aca}
    M.~G.~Alford, S.~Han and M.~Prakash,
    Phys.\ Rev.\ D {\bf 88} (2013) no.8,  083013.
  
\bibitem{Glendenning:1998ag}
  N.~K.~Glendenning and C.~Kettner,
  Astron.\ Astrophys.\  {\bf 353} (2000) L9.

\bibitem{Blaschke:2013ana}
  D.~Blaschke, D.~E.~Alvarez-Castillo and S.~Benic,
  PoS  {\bf CPOD 2013} (2013) 063.

\bibitem{Benic:2014jia}
   S.~Benic, D.~Blaschke, D.~E.~Alvarez-Castillo, T.~Fischer and S.~Typel,
   Astron.\ Astrophys.\  {\bf 577} (2015) A40.
  
\bibitem{Alvarez-Castillo:2017qki}
    D.~E.~Alvarez-Castillo and D.~B.~Blaschke,
    Phys.\ Rev.\ C {\bf 96} (2017) no.4,  045809.


\bibitem{Owen:1977uun}
  J.~C.~Owen, R.~F.~Bishop and J.~M.~Irvine,
  Nucl.\ Phys.\ A {\bf 277} (1977) 45.
    \bibitem{Modarres:1979jk}
        M.~Modarres and J.~M.~Irvine,
        J.\ Phys.\ G {\bf 5} (1979) 511.
     \bibitem{Bordbar:1998xv}
         G.~H.~Bordbar and M.~Modarres,
         Phys.\ Rev.\ C {\bf 57} (1998) 714.
      \bibitem{Modarres:1998ma}
        M.~Modarres and G.~H.~Bordbar,
        Phys.\ Rev.\ C {\bf 58} (1998) 2781.
       \bibitem{Modarres:2000nk}
         M.~Modarres and H.~R.~Moshfegh,
         Phys.\ Rev.\ C {\bf 62} (2000) 044308.
       \bibitem{Moshfegh:2005rom}
         H.~R.~Moshfegh and M.~Modarres,
         Nucl.\ Phys.\ A {\bf 759} (2005) 79.
         
        \bibitem{Moshfegh:2007mxh}
           H.~R.~Moshfegh and M.~Modarres,
           Nucl.\ Phys.\ A {\bf 792} (2007) 201.
      \bibitem{Hiyama:2006xv}
      E.~Hiyama, Y.~Yamamoto, T.~A.~Rijken and T.~Motoba,
      Phys.\ Rev.\ C {\bf 74} (2006) 054312.
    \bibitem{Hiyama:2002yj}
      E.~Hiyama, M.~Kamimura, T.~Motoba, T.~Yamada and Y.~Yamamoto,
      Phys.\ Rev.\ C {\bf 66} (2002) 024007.

\bibitem{Goudarzi:2015dax}
  S.~Goudarzi and H.~R.~Moshfegh,
  Phys.\ Rev.\ C {\bf 91} (2015) no.5,  054320,
   Erratum: [Phys.\ Rev.\ C {\bf 96} (2017) no.4,  049902].

   \bibitem{Goudarzi:2019orb} 
  S.~Goudarzi and H.~R.~Moshfegh,
  Nucl.\ Phys.\ A {\bf 985}, (2019) 1.
 
\bibitem{modarres2009} 
M. Modarres and H. R. Moshfegh, Physica A, {\bf 388} (2009) 3297.

\bibitem{modarres2003} 
M. Modarres, H. R. Moshfegh, and A. Sepahvand, Eur. Phys. J. B, {\bf 31} (2003) 159. 

\bibitem{Akmal:1998cf}
  A.~Akmal, V.~R.~Pandharipande and D.~G.~Ravenhall,
  Phys.\ Rev.\ C {\bf 58} (1998) 1804.

\bibitem{moshfegh2010} 
H. R. Moshfegh and S. Zaryouni, Eur. Phys. J. A {\bf 43} (2010) 283; {\bf 45} (2010) 69.     

\bibitem{Wiringa:1994wb}
  R.~B.~Wiringa, V.~G.~J.~Stoks and R.~Schiavilla,
  Phys.\ Rev.\ C {\bf 51} (1995) 38.
  

\bibitem{Masuda:2012ed}
  K.~Masuda, T.~Hatsuda and T.~Takatsuka,
  PTEP {\bf 2013} (2013) no.7,  073D01.

\bibitem{Asakawa:1995zu}
  M.~Asakawa and T.~Hatsuda,
  Phys.\ Rev.\ D {\bf 55} (1997) 4488.

\bibitem{Blaschke:2013rma}
  D.~Blaschke, D.~E.~Alvarez Castillo, S.~Benic, G.~Contrera and R.~Lastowiecki,
  PoS {\bf  ConfinementX} (2012) 249.
  
\bibitem{Zdunik:2012dj}
  J.~L.~Zdunik and P.~Haensel,
  Astron.\ Astrophys.\  {\bf 551} (2013) A61.

\bibitem{Alford:2015gna}
  M.~G.~Alford and S.~Han,
  Eur.\ Phys.\ J.\ A {\bf 52} (2016) no.3,  62.

\bibitem{Paschalidis:2017qmb}
  V.~Paschalidis, K.~Yagi, D.~Alvarez-Castillo, D.~B.~Blaschke and A.~Sedrakian,
  Phys.\ Rev.\ D {\bf 97} (2018) no.8,  084038.

\bibitem{Alford:2017qgh}
  M.~G.~Alford and A.~Sedrakian,
  Phys.\ Rev.\ Lett.\  {\bf 119} (2017) no.16,  161104.

\bibitem{Tolman:1939jz}
  R.~C.~Tolman,
  Phys.\ Rev.\  {\bf 55} (1939) 364.

\bibitem{Oppenheimer:1939ne}
  J.~R.~Oppenheimer and G.~M.~Volkoff,
  Phys.\ Rev.\  {\bf 55} (1939) 374.

\bibitem{Negele:1971vb}
  J.~W.~Negele and D.~Vautherin,
  Nucl.\ Phys.\ A {\bf 207} (1973) 298.

\bibitem{harrisonwheeler} 
B. K. Harrison, K. S. Thorne, M. Wakano, J. A. Wheeler, 
Gravitation Theory and Gravitational Collapse, University of Chicago Press, Chicago, 1965. 

\bibitem{Bauswein:2018bma}
  A.~Bauswein, N.~U.~F.~Bastian, D.~B.~Blaschke, K.~Chatziioannou, J.~A.~Clark, T.~Fischer and M.~Oertel,
  Phys.\ Rev.\ Lett.\  {\bf 122} (2019) no.6,  061102.

\bibitem{Fischer:2017lag}
  T.~Fischer {\it et al.},
  Nat.\ Astron.\  {\bf 2} (2018) no.12,  980.

\end{thebibliography}

\newpage

\begin{figure*}
\centering
\includegraphics[width=\textwidth]{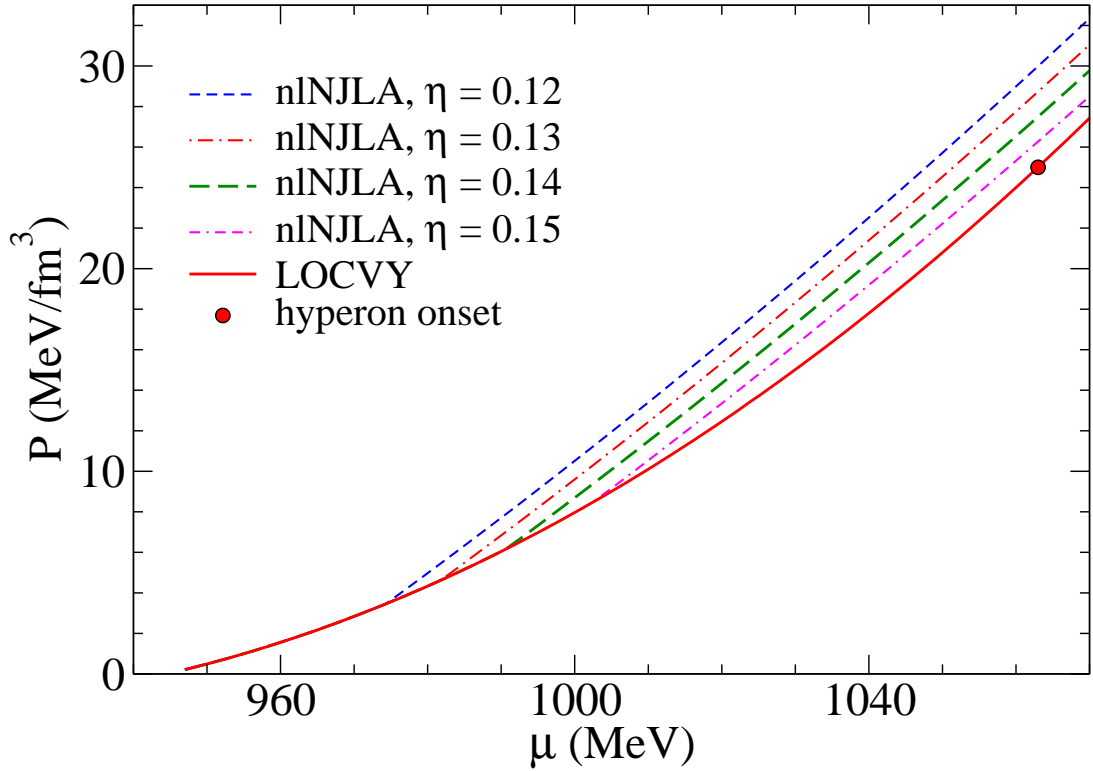}
\caption{\label{fig:2} Pressure as a function of chemical potential for the Maxwell construction of the deconfinement PT using the LOCV method with hyperons (LOCVY) for the hadronic phase and the color superconducting nlNJLA model for quark matter. The onset of deconfinement can be constructed for parametrisations with $\eta< 0.16$ and occurs before the onset of hyperons.}
\end{figure*}
\begin{figure*}
\centering
\includegraphics[width=\textwidth]{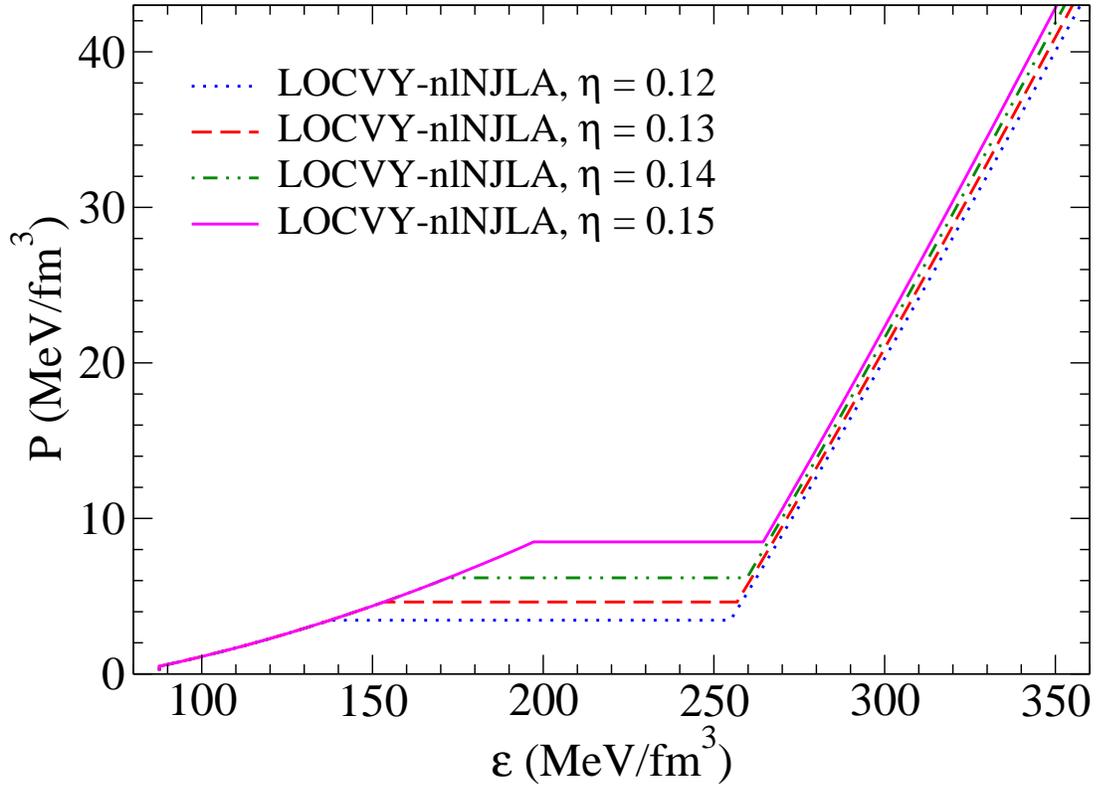}
\caption{\label{fig:3} Pressure as a function of energy density for Maxwell construction of the deconfinement PT using the LOCV method with hyperons (LOCVY) for the hadronic phase and the color superconducting nlNJLA model for quark matter.}
\end{figure*}
\begin{figure*}
\centering
\includegraphics[width=\textwidth]{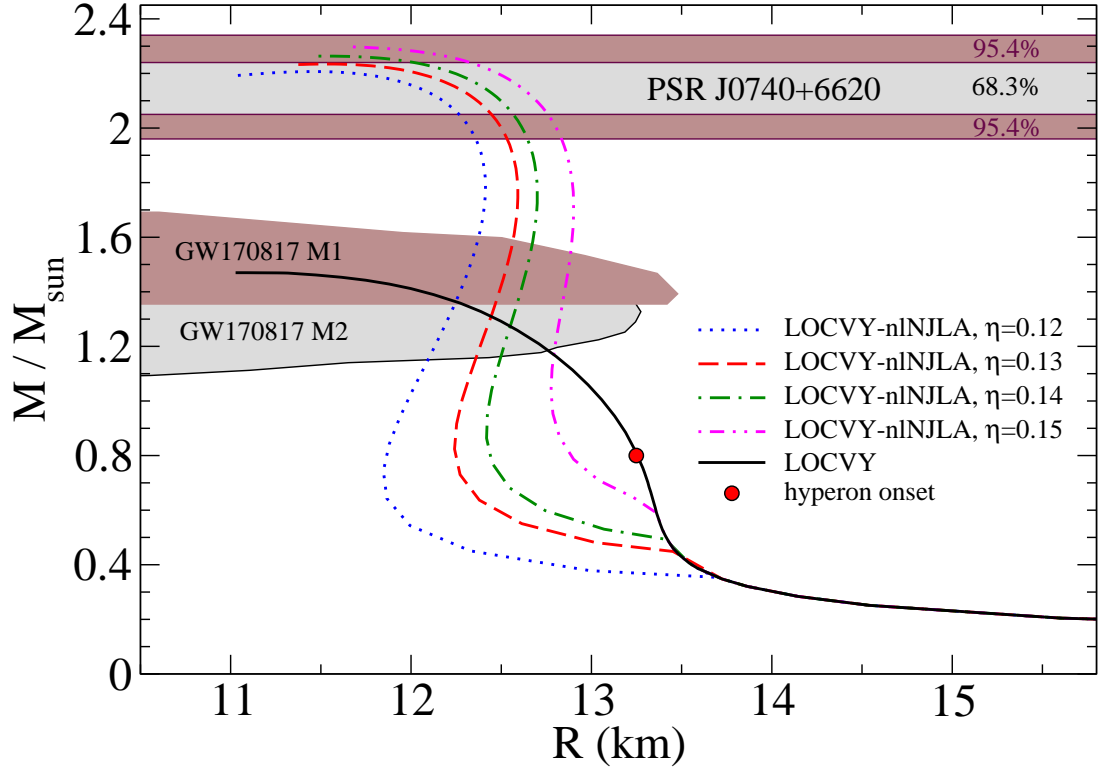}
\caption{\label{fig:4} Mass-radius relation of hybrid star for Maxwell construction of the deconfinement PT using LOCV method with hyperons (LOCVY) for the hadronic phase and the color superconducting nlNJLA model for quark matter.
The new observational constraint for the maximum mass from PSR J0740+6620 \cite{Cromartie:2019kug} and compactness constraint from the LIGO analysis of GW170817 \cite{TheLIGOScientific:2017qsa}. 
}
\end{figure*}
\begin{figure*}
\centering
\includegraphics[width=\textwidth]{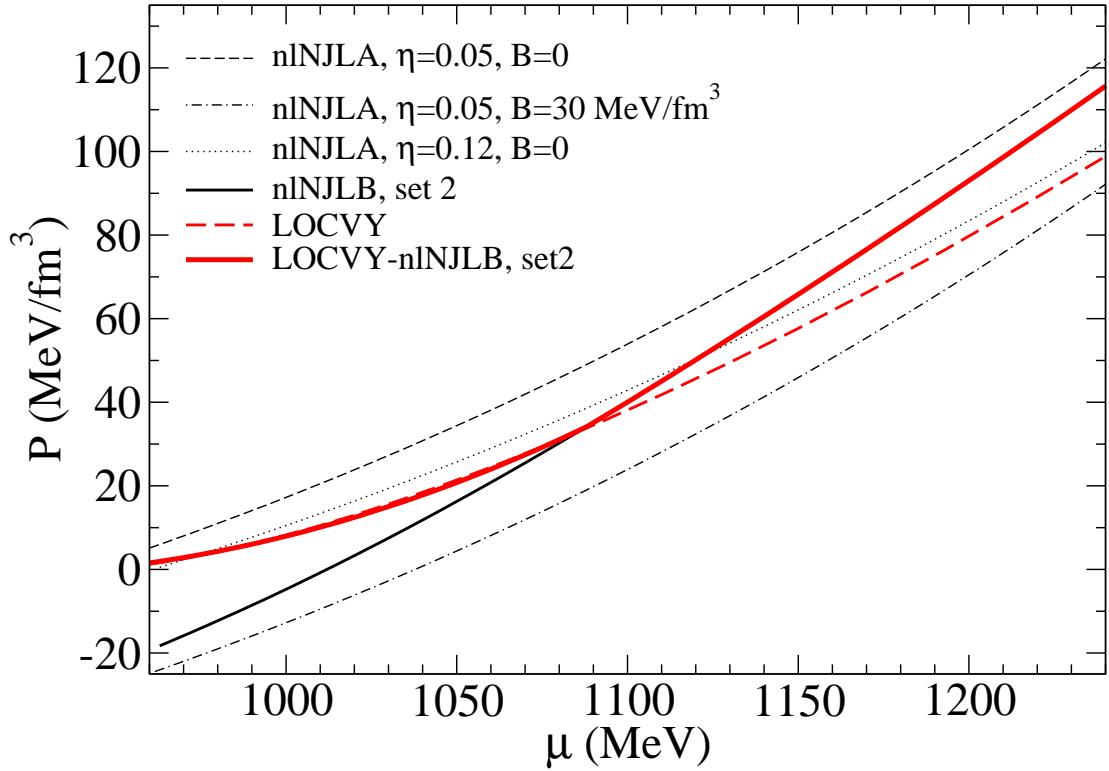}
\caption{\label{fig:5} 
Hybrid star EoS (bold solid line) obtained by a Maxwell construction between the nlNJLB quark matter EoS with density dependent coefficients according to set 2 and the LOCVY EoS for hypernuclear matter in $\beta$-equilibrium with electrons, muons as well as $\Sigma^-$ and $\Lambda$ hyperons. 
The quark matter EoS is based on three parametrizations of the nlNJLA model with constant coefficients: a soft (low $\eta$) one with confinement $(B\ne0)$ at low densities, a soft one without confinement $B=0$ at intermediate densities and a stiff one (high $\eta$) at high densities.}
\end{figure*}
\begin{figure*}
\centering
\includegraphics[width=\textwidth]{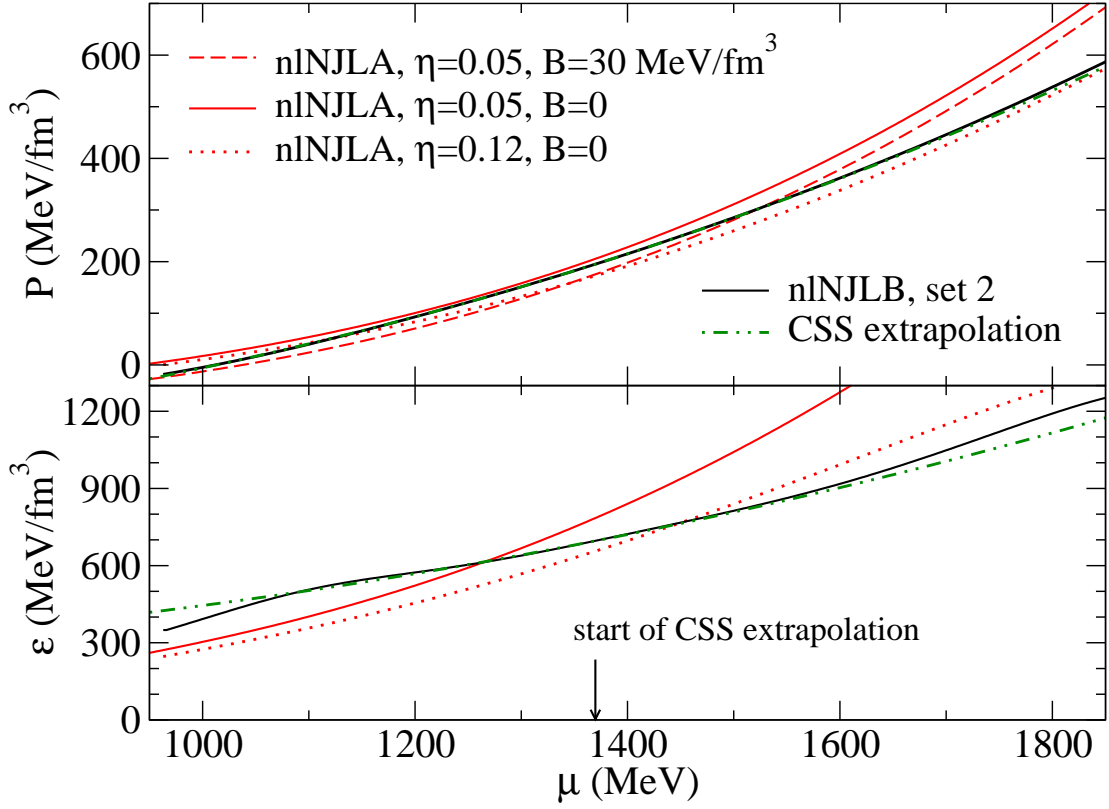}
\caption{\label{fig:6} Quark matter EoS for which we have used the nlNJLB quark matter model with density dependent coefficients according to set 2 at low and intermediate chemical potential while for high chemical potential, the CSS extrapolation method has been used. The matching point is at $\varepsilon=690$ MeV/fm$^3$ which for nlNJLB set 2 corresponds to $\mu=1362$ MeV. 
The top panel shows the pressure as a function of the chemical potential in which there is a good agreement between the nlNJLB EoS and extrapolated one. The lower plot shows the energy density as a function of chemical potential.}
\end{figure*}
\begin{figure*}
\centering
\includegraphics[width=\textwidth]{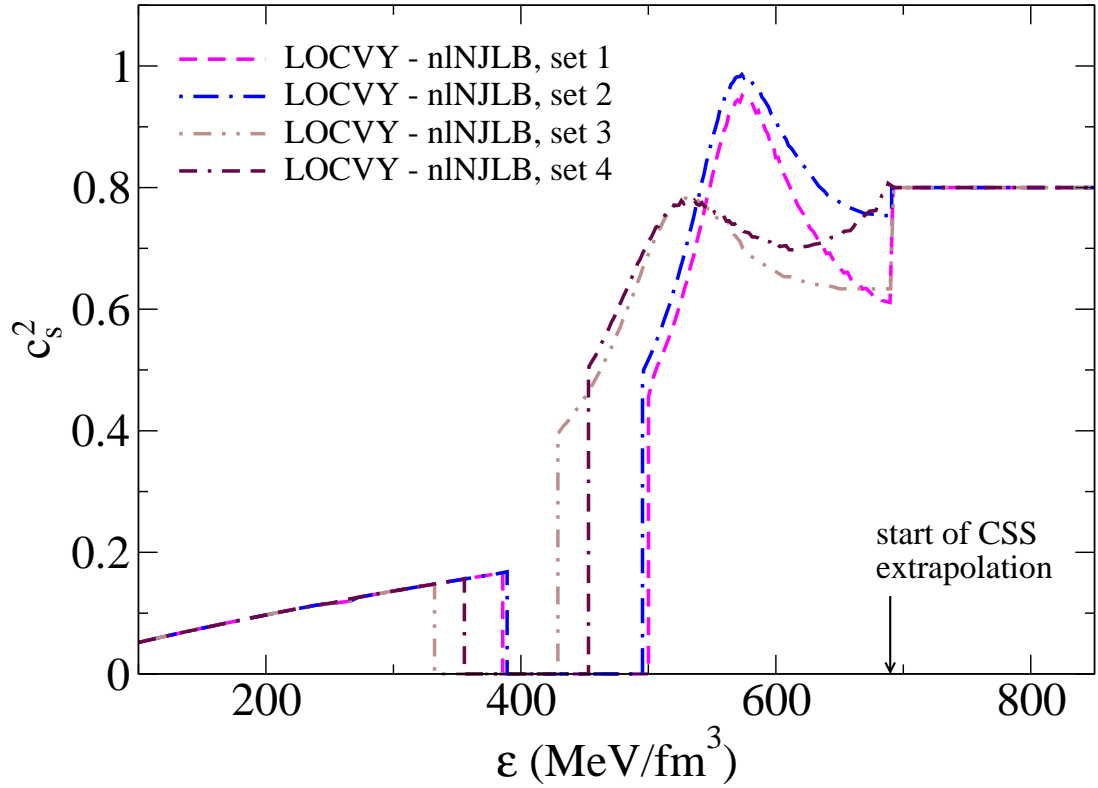}
\caption{\label{fig:7} The squared speed of sound $c_s^2$ in units of the squared speed of light as a function of the energy density for all sets in Table I.
The region of high energy densities ($>690$ MeV/fm$^3$) is described by a constant speed of sound model.}
\end{figure*}
\begin{figure*}
\centering
\includegraphics[width=\textwidth]{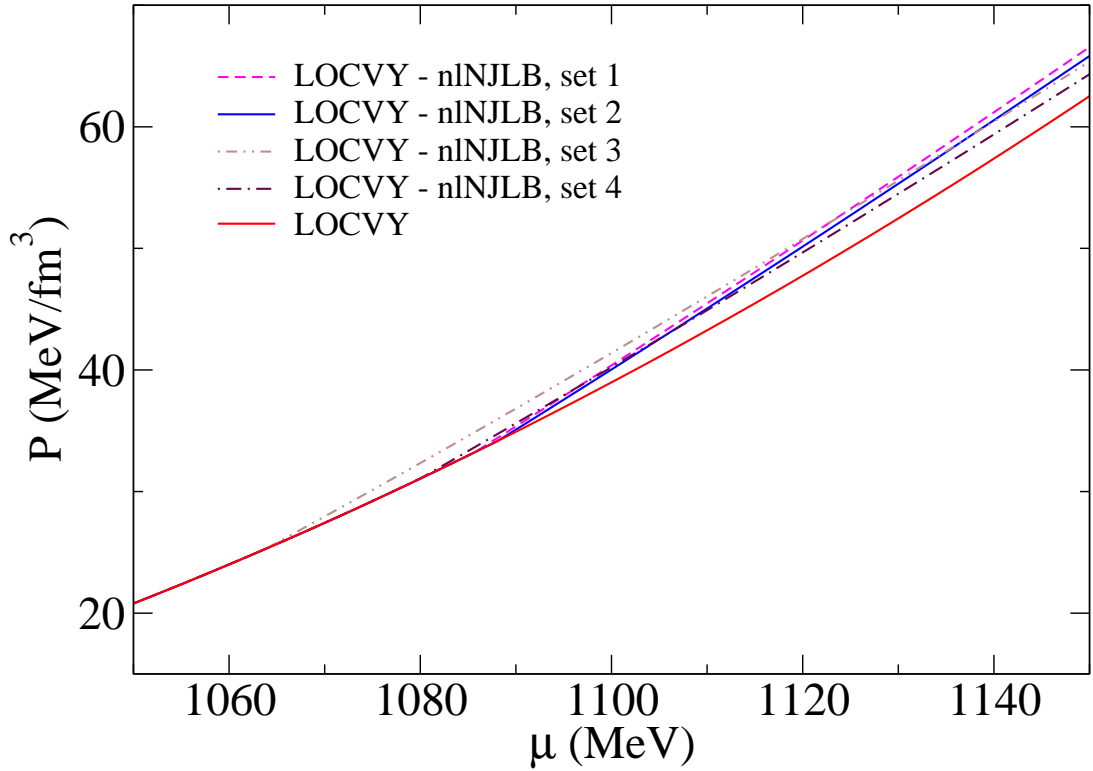}
\caption{\label{:wide.} Pressure as a function of chemical potential for the Maxwell construction of the deconfinement PT using the LOCV method for hypernuclear matter (LOCVY) and four sets of parametrization of the model nlNJLB for quark matter. The EoS of pure hypernuclear matter is shown as well.}
\end{figure*}
\begin{figure*}
\centering
\includegraphics[width=\textwidth]{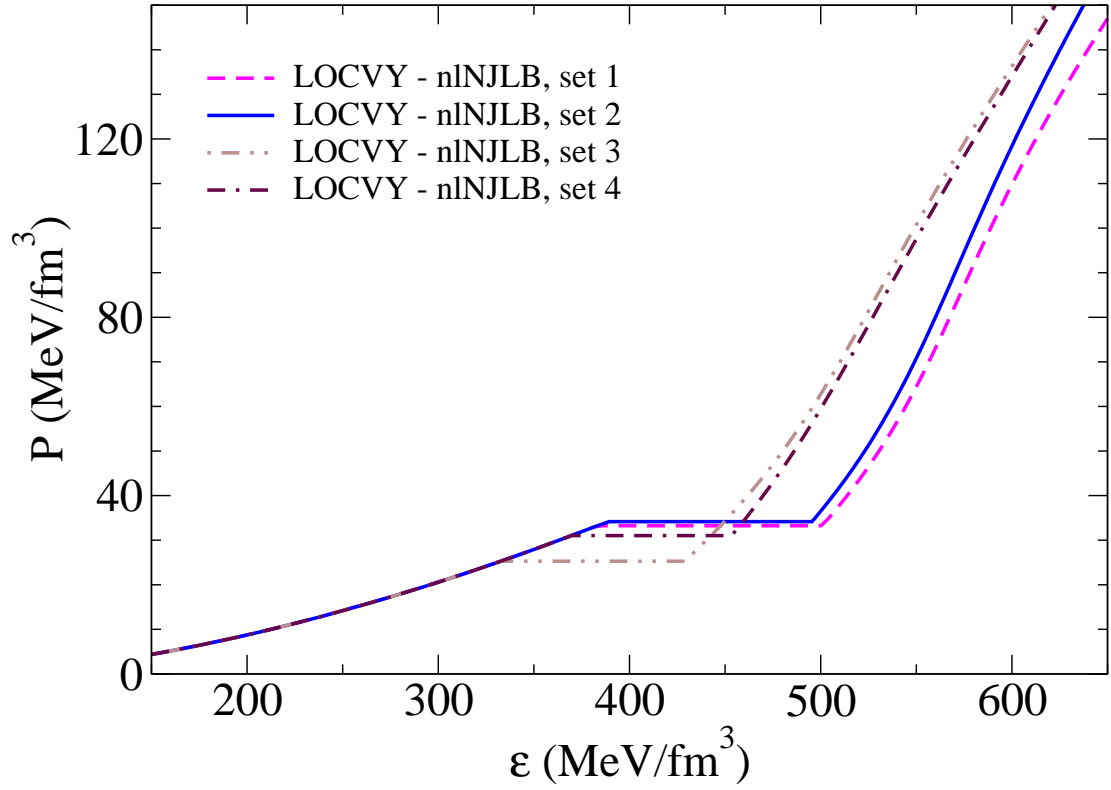}
\caption{\label{fig:8} Pressure as a function of energy density for the Maxwell construction of the deconfinement PT using LOCV method for hypernuclear matter (LOCVY) and four sets of parametrization of the model nlNJLB for quark matter.}
\end{figure*}
\begin{figure*}
\centering
\includegraphics[width=\textwidth]{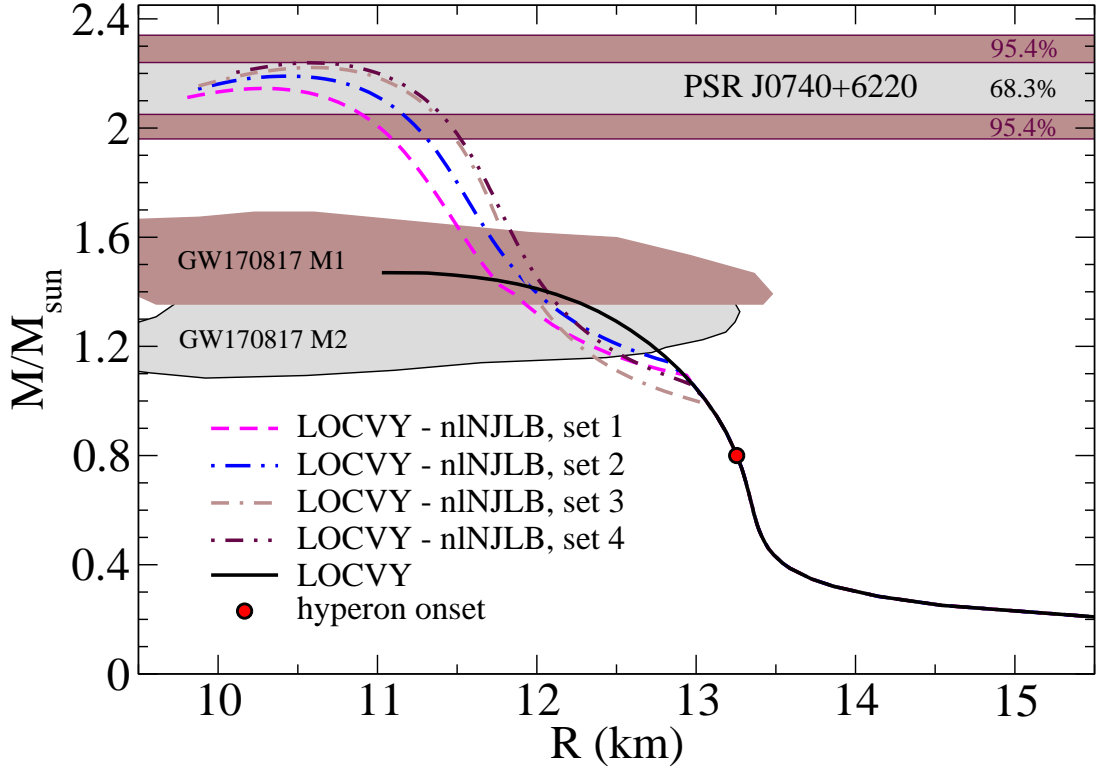}
\caption{\label{fig:9} Mass-radius relation of hybrid star for Maxwell construction of the deconfinement PT using LOCV method for hypernuclear matter (LOCVY) and four sets of parametrization of the nlNJLB  model for quark matter. The hyperon onset (filled circle) occurs at a star mass of $0.8~M_\odot$, before the quark deconfinement, so that above that mass there are hybrid stars with three phases of core matter: nuclear, hypernuclear and quark matter, see Figs.~\ref{fig:profiles-low} and \ref{fig:profiles-high}.}
\end{figure*}

\begin{figure*}
\includegraphics[width=0.7\textwidth]{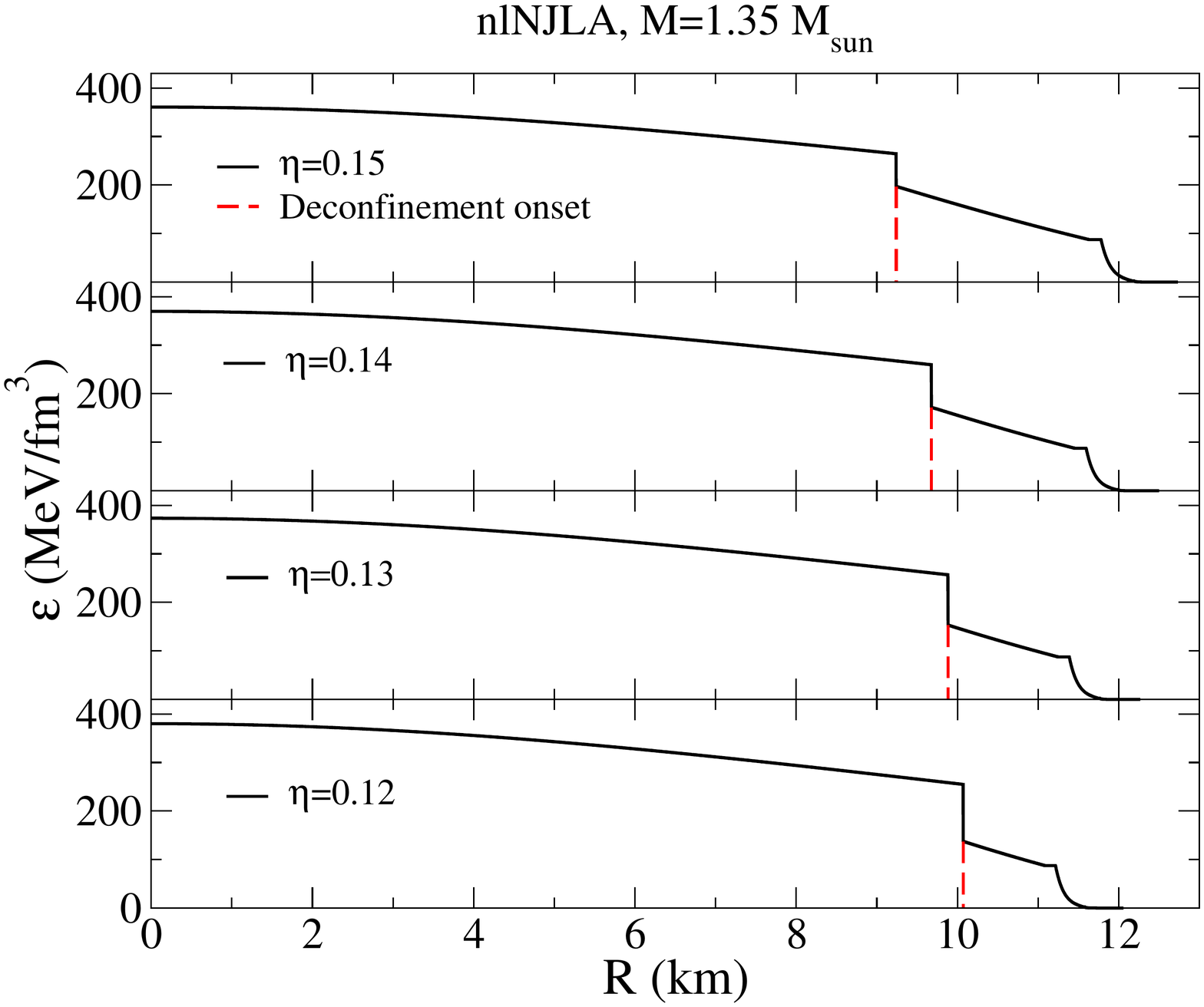}\\
\includegraphics[width=0.7\textwidth]{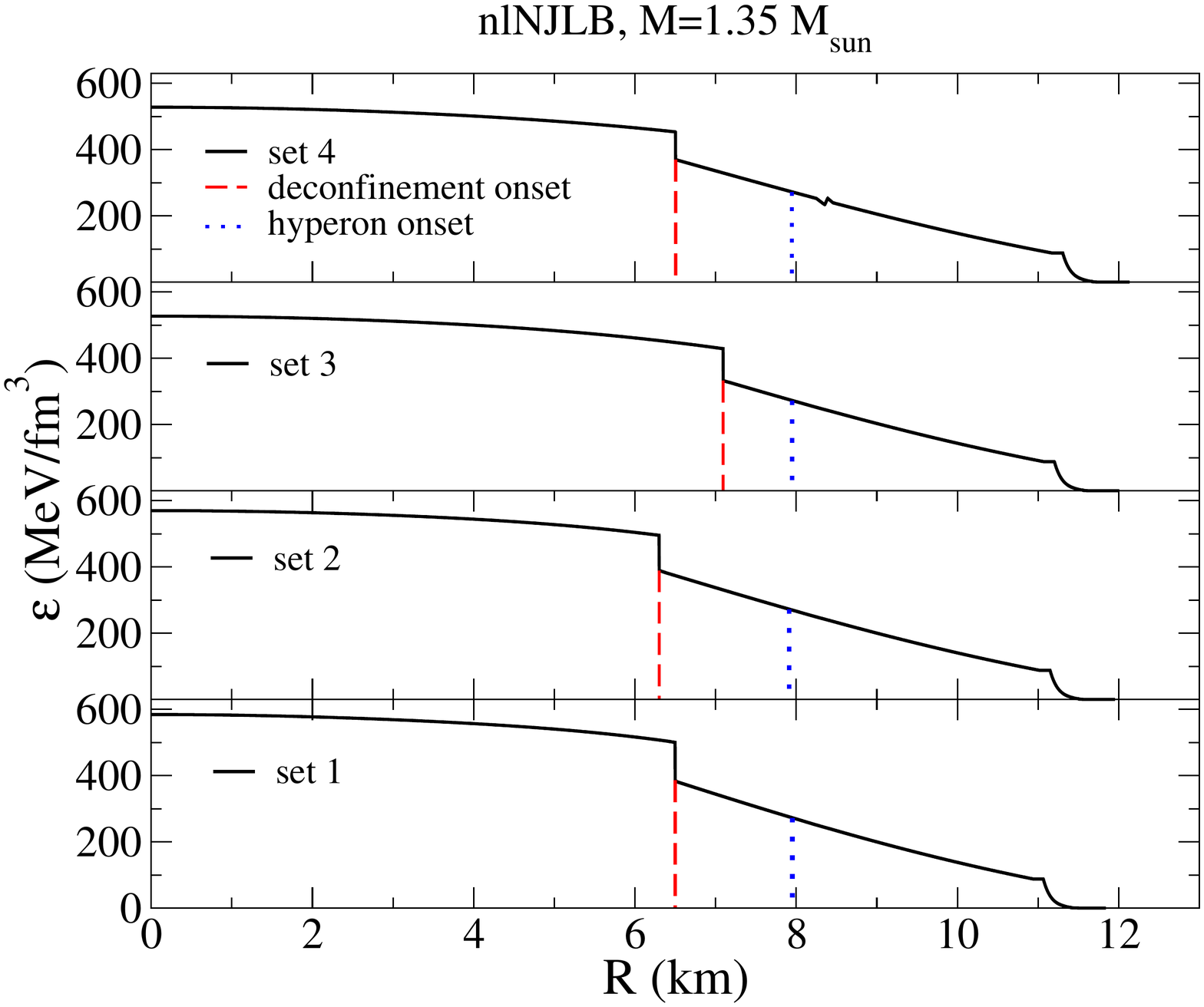}
\caption{\label{fig:profiles-low} Profiles of energy densities for model nlNJLA (upper panel) with four cases of vector coupling strength
$\eta=0.12, 0.13, 0.14, 0.15$ and the four sets of parametrisation of model nlNJLB (lower panel) for the case of a typical binary radio pulsar mass of $1.35~M_\odot$.
While for the parametrizations of model nlNJLA the deconfinement transition occurs directly from the nuclear matter outer core to the extended quark matter core, in model nlNJLB there is a shell of hypernuclear matter in-between the inner core of color superconducting  quark matter and the outer core of nuclear matter.
}
\end{figure*}

\begin{figure*}
\includegraphics[width=0.7\textwidth]{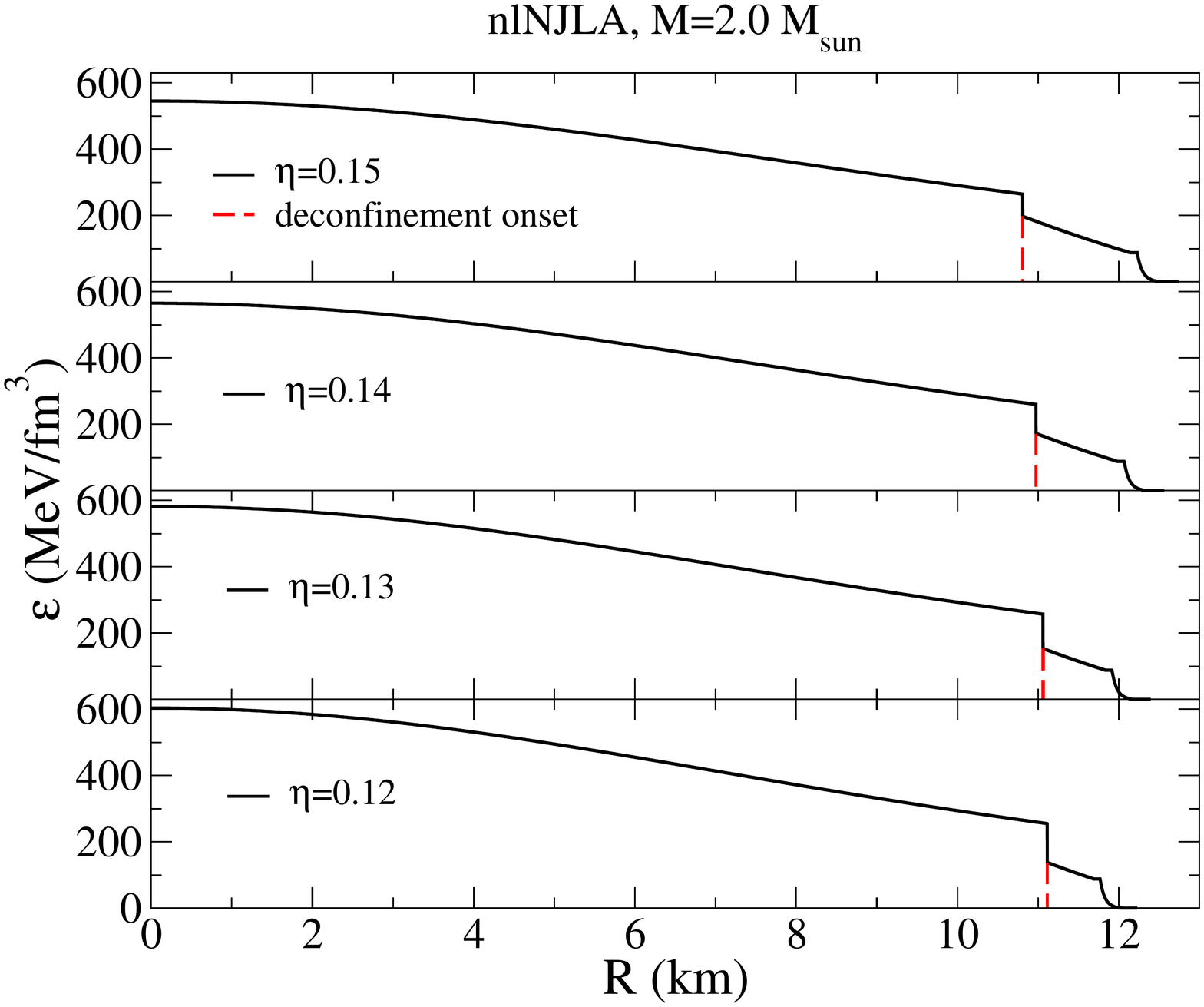}\\
\includegraphics[width=0.7\textwidth]{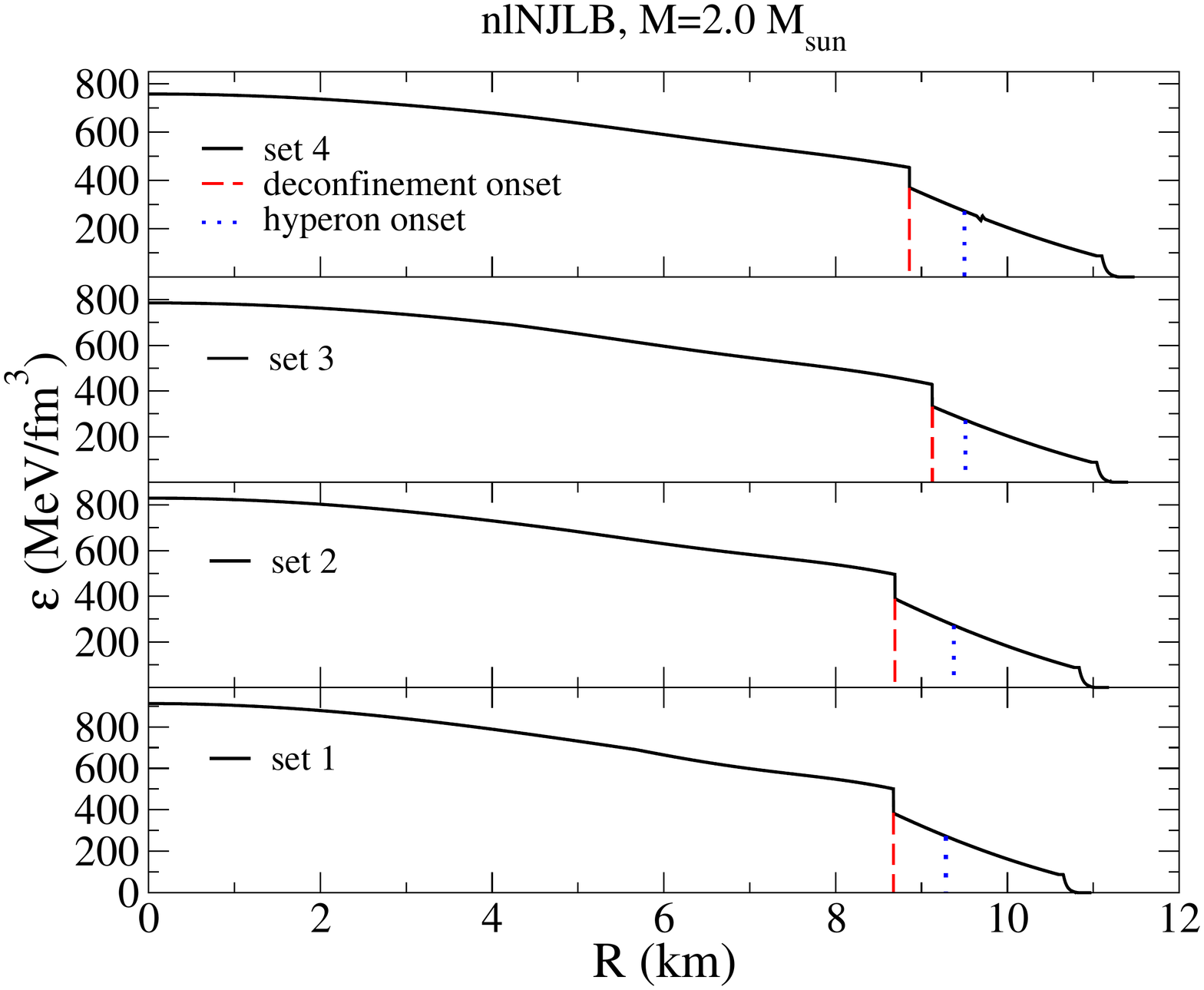}
\caption{\label{fig:profiles-high} Profiles of energy densities for model nlNJLA (upper panel) with four cases of vector coupling strength
$\eta=0.12, 0.13, 0.14, 0.15$ and the four sets of parametrisation of model nlNJLB (lower panel) for the case of a high-mass 
pulsar with $2.0~M_\odot$.
While for parametrizations of model nlNJLA the deconfinement transition occurs directly from the nuclear matter outer core to the extended quark matter core, in model nlNJLB there is a shell of hypernuclear matter in-between the inner core of color superconducting  quark matter and the outer core of nuclear matter.
}
\end{figure*}

\begin{figure*}
\includegraphics[width=0.65\textwidth]{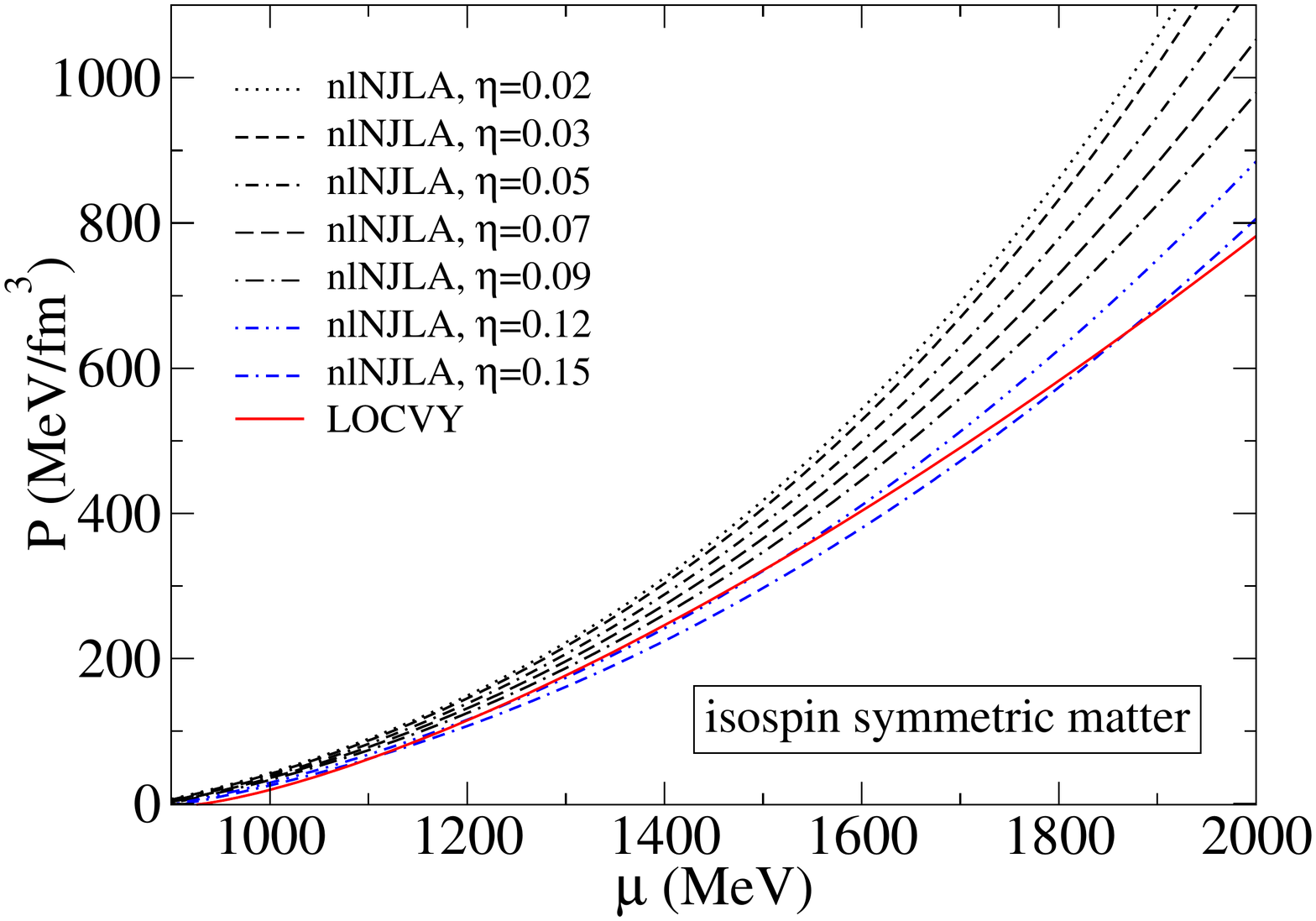}
\vspace{-1cm}

\includegraphics[width=0.65\textwidth]{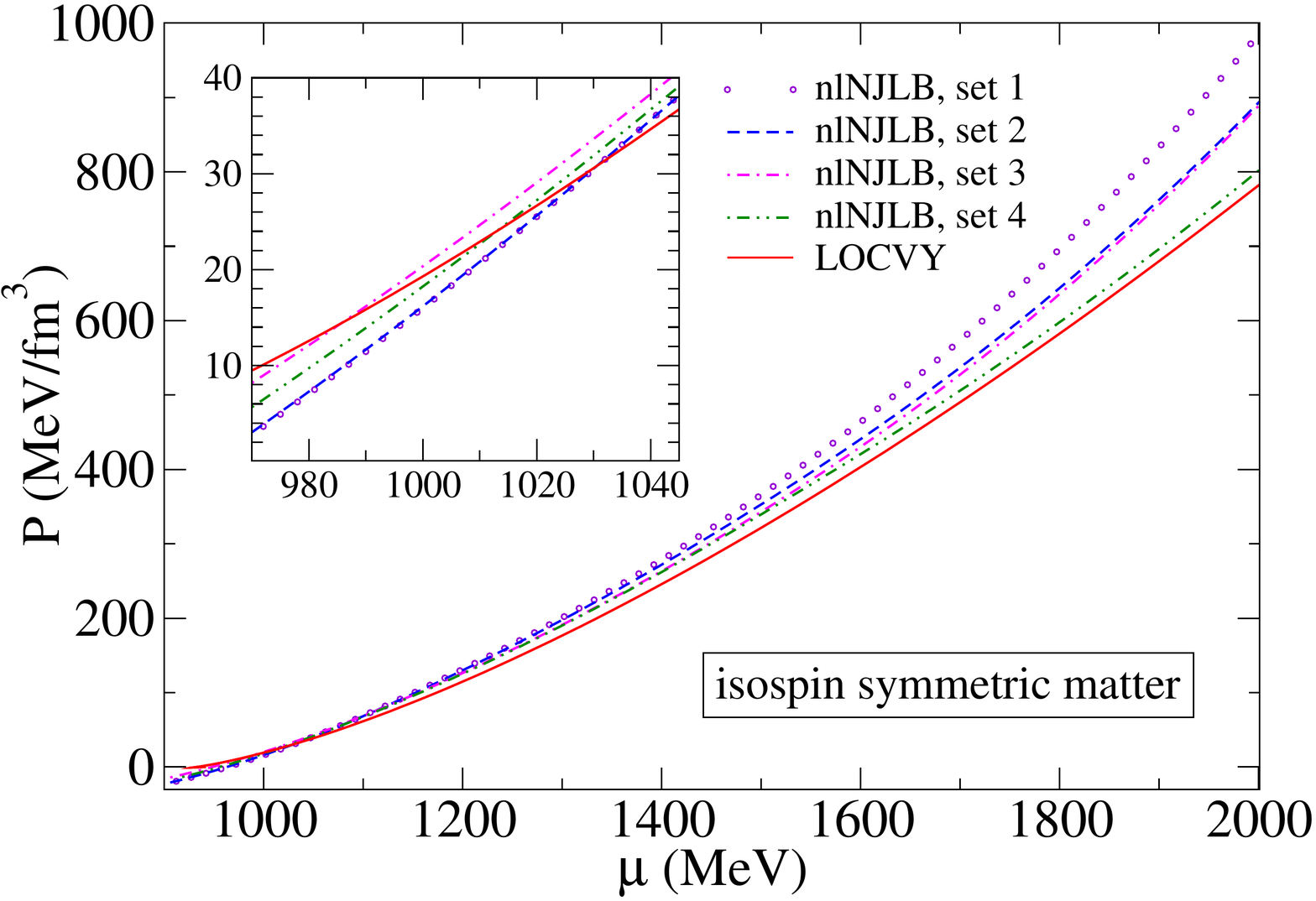}
\vspace{-1cm}

\caption{\label{fig:PT_sym} Pressure versus baryon chemical potential in the isospin-symmetric case for the LOCVY model of hypernuclear matter and for the nlNJL approach to quark matter. 
Upper panel: model nlNJLA with vector coupling strengths $\eta=0.02, 0.03, 0.05, 0.07, 0.09, 0.12, 0.15$; 
lower panel: model nlNJLB for sets 1 - 4. 
For model nlNJLA there is obviously a problem at low chemical potential, where due to the lack of confinement in this model the quark pressure is above the hadronic one. For the cases with $\eta > 0.12$ (which produce reasonable hybrid star EoS) there are also reasonable phase transitions when one ignores the region below $\mu\approx 1400$ MeV. 
For $\eta = 0.12$ the critical density for the onset of deconfinement is $n_c=0.79$ fm$^{-3}$ and for $\eta = 0.15$ it is $n_c=0.98$ fm$^{-3}$.
For model nlNJLB under isospin-symmetric conditions all considered parametrizations of sets 1 - 4 predict a deconfinement transition between 2.2 $n_0$ and 2.7  $n_0$.
For a detailed discussion, see text.
}
\end{figure*}

\end{document}